\documentclass[pra,aps,twocolumn,superscriptaddress,showpacs,10pt,floatfix]{revtex4}

\usepackage{amsmath}
\usepackage{amssymb}
\usepackage{graphicx}
\usepackage{microtype}

\usepackage[svgnames]{xcolor}
\usepackage[pdftex]{hyperref}
\hypersetup{colorlinks=true,linkcolor=blue,citecolor=ForestGreen,urlcolor=DarkOrchid}

\graphicspath{{images/}}

\renewcommand{\appendixname}{Appendix}
\newcommand{\equationname}{Equation}
\newcommand{\referencename}{Ref.}
\renewcommand{\figurename}{Figure}
\renewcommand{\tablename}{Table}
\newcommand{\kff}{$\text{K}_{4,4}$ }

\def\ket#1{\left|#1\right\rangle}

\def\ketbra#1#2{\left|#1\rangle\!\langle#2\right|}

\def\DW{\mbox{DW2X$\ $}}

\newcommand{\harvard}{\affiliation{Department of Chemistry and Chemical Biology, Harvard 
University, 12 Oxford Street, 02138 Cambridge, Massachusetts, USA}}
\newcommand{\tamu}{\affiliation{Department of Physics and Astronomy, Texas A\&M University, 
College Station, Texas 77843-4242, USA}}
\newcommand{\nasa}{\affiliation{Quantum Artificial Intelligence Lab., NASA Ames Research Center, Moffett Field, CA 94035, USA}}
\newcommand{\ucscn}{\affiliation{University of California Santa Cruz @ NASA Ames Research Center, Moffett Field, CA 94035, USA}}
\newcommand{\sfi}{\affiliation{Santa Fe Institute, 1399 Hyde Park Road, Santa Fe, New Mexico 
87501 USA}}
\newcommand{\coventry}{\affiliation{Applied Mathematics Research Centre, Coventry University, 
Coventry, CV1 5FB, United Kingdom}}

\begin{document}

\title{Strengths and weaknesses of weak-strong cluster problems:\\ A
detailed overview of state-of-the-art classical heuristics vs quantum
approaches}

\author{Salvatore Mandr{\`a}}
\email{smandra@fas.harvard.edu}
\harvard

\author{Zheng Zhu}
\email{zzwtgts@tamu.edu}
\tamu

\author{Wenlong Wang}
\email{wenlong@physics.umass.edu}
\tamu

\author{Alejandro Perdomo-Ortiz}
\email{alejandro.perdomoortiz@nasa.gov}
\nasa 
\ucscn

\author{Helmut G.~Katzgraber}
\email{hgk@tamu.edu}
\tamu
\sfi
\coventry

\date{\today}

\begin{abstract}

To date, a conclusive detection of quantum speedup remains elusive.
Recently, a team by Google Inc.~[V.~S.~Denchev {\em et al}.,
Phys.~Rev.~X {\bf 6}, 031015 (2016)] proposed a weak-strong cluster
model tailored to have tall and narrow energy barriers separating local
minima, with the aim to highlight the value of finite-range tunneling.
More precisely, results from quantum Monte Carlo simulations, as well as
the D-Wave 2X quantum annealer scale considerably better than
state-of-the-art simulated annealing simulations. Moreover, the D-Wave
2X quantum annealer is $\sim 10^8$ times faster than simulated annealing
on conventional computer hardware for problems with approximately $10^3$
variables. Here, an overview of different sequential, nontailored, as
well as specialized tailored algorithms on the Google instances is
given. We show that the quantum speedup is limited to sequential
approaches and study the typical complexity of the benchmark problems
using insights from the study of spin glasses.

\end{abstract}


\maketitle

\section{Introduction}
\label{sec:intro}

Adiabatic quantum optimization (QA)
\cite{nielsen:00,nishimori:01,finnila:94,kadowaki:98,brooke:99,farhi:00,roland:02,santoro:02,das:05,santoro:06,lidar:08,das:08,morita:08,mukherjee:15},
the quantum version of classical simulated annealing (SA)
\cite{kirkpatrick:83}, has caused considerable controversy and interest
since the introduction of the D-Wave Inc.~\cite{comment:d-wave} quantum
annealing machines \cite{johnson:11}.  Although there is increasing
evidence that quantum effects do play a role in the optimization process
of these machines, there is still no consensus as to if the machine is
able to outperform classical optimization heuristics on silicon-based
computer hardware. Multiple teams
\cite{dickson:13,pudenz:13,smith:13,boixo:13a,ronnow:14,katzgraber:14,lanting:14,santra:14,shin:14,boixo:14,albash:15,albash:15a,katzgraber:15,martin-mayor:15,pudenz:15,hen:15a,venturelli:15a,vinci:15,zhu:16}
have scrutinized this first commercially available programmable analog
quantum optimizer [the current version being the D-Wave 2X (DW2X) with
up to $1152$ quantum bits wired on a chimera topology \cite{bunyk:14}]
and tried to understand its advantages and disadvantages over classical
technologies, as well as improve its performance via, e.g., quantum
error correction \cite{lidar:13,pudenz:15,vinci:15} (at the price of
having too few logical qubits for a scaling analysis) or fine-tuning of
the device \cite{perdomo:15,perdomo:15a}.

In an effort to determine the thermodynamic (large number of qubits $n$)
scaling advantage of a quantum annealer over conventional algorithms, it
is of importance to use the largest possible number of qubits on any
device. As such, embedded problems (that might require an overhead due
to the embedding) are sub-optimal for scaling analyses. {\em Native}
problems, such as spin-glass-like systems \cite{binder:86,stein:13} that
use all qubits of the system, are thus optimal to tickle out any
putative quantum advantage from quantum annealing machines.
Unfortunately, results have been inconclusive so far \cite{ronnow:14}
and there is strong evidence that random spin-glass problems are not
well suited for benchmarking purposes
\cite{katzgraber:14,katzgraber:15}. Thus, efforts have shifted to {\em
tailored} problems, such as carefully-crafted spin-glass instances
\cite{katzgraber:15,zhu:16} that are robust to the intrinsic noise of
analog machines. In particular, \referencename~\cite{katzgraber:15}
suggested a slight quantum advantage over classical simulated annealing
\cite{isakov:15,isakov:14a} using the $512$-qubit D-Wave 2 quantum
annealer \cite{comment:relaxed}.  However, no scaling analysis was
performed because systems of approximately $n \sim 500$ qubits are just
at the brink of the scaling regime.

Despite all these efforts, a ``killer'' application or problem domain
has yet to be found, where quantum annealing outperforms notably
classical simulational approaches.  In particular, given that many
well-known optimization problems from the traveling salesman problem to
constraint-satisfaction and vertex cover problems can be mapped onto
Ising spin-glass-like Hamiltonians \cite{lucas:14}, there is great
interest from both science and industry to find efficient optimization
approaches to tackle spin-glass-like Hamiltonians -- the main forte of
the \DW device.  Most recently, however, a team by Google
Inc.~\cite{denchev:16} showed for carefully-crafted problems that
quantum annealing on the \DW can outperform classical simulated
annealing by approximately eight orders of magnitude. Furthermore, the
scaling of quantum approaches (both on the \DW and using quantum Monte
Carlo \cite{suzuki:93}) is considerably better than for classical
simulated annealing. We believe this is the first notable ``success
story'' for quantum annealing.  However, numerical comparisons were
performed against one of the commonly-known least-efficient optimization
methods, namely simulated annealing. While this seems to be a fair
comparison because both QA and SA are sequential optimization methods
where a control parameter (quantum fluctuations in the former and
thermal fluctuations in the latter) is decreased monotonically until
reaching a target value, it is unclear if this favorable scaling will
persist for state-of-the art optimization methods (see, for example,
Refs.~\cite{hartmann:01,juenger:01,hartmann:04,zhu:15b} for some
examples). We do emphasize, however, that the Google Inc.~studies
\cite{boixo:14a,denchev:16,boixo:16} shed, for the first time, some
light on the structure of problems where quantum annealing might excel.
In particular, by carefully crafting weak-strong cluster problems (see
Sec.~\ref{sec:wscm} for details), they can show that there is a
sign of finite-range quantum tunneling, at least within the basic
building blocks of the \DW device, known as a \kff cell
\cite{denchev:16}.

In this work, we complement the study of
Ref.~\cite{denchev:16}: first, we expand the notion of
``limited quantum speedup'' \cite{ronnow:14} to take into account
different classes of algorithms (see Sec.~\ref{sec:qs}) and thus
attempt to present a fair assessment of any sequential quantum annealer.
In particular, we introduce the notion of ``{\em limited sequential
quantum speedup}'' which refers to speedup with respect to any algorithm
that optimizes sequentially such as, for example, simulated annealing.
Furthermore, we distinguish two types of state-of-the-art optimization
methods: ``tailored'' and ``nontailored'' algorithms. Tailored
algorithms exploit the structure of the studied optimization problem; we
thus feel this might pose an unfair advantage. Nontailored algorithms
are generic, and thus present the state of the art when studying a wide
variety of optimization problems.  Our results show that sequential
quantum approaches (\DW quantum annealer and quantum Monte Carlo)
clearly outperform any other currently-available sequential methods, but
fall short of outperforming nontailored (as well as tailored)
algorithms. We thus herewith raise the bar for any quantum optimization
approach.  Second, we illustrate with a simple two-energy level model
with noise, how a suboptimal annealing time for small problem sizes can
lead to a change in slope of the scaling analysis, as observed in
Ref.~\cite{denchev:16} for the \DW machine. Finally, we study the energy
landscape of the instances and show that the spin-glass backbone of the
weak-strong cluster network dominates and thus might negatively impact
the scaling of this class of problem for future larger chips and/or
system sizes.

The paper is structured as follows. In Sec.~\ref{sec:qs}, we
introduce a classification for the concept of ``quantum speedup'',
in order to better assess the comparison between classical and quantum
devices.  In Sec.~\ref{sec:wscm}, we briefly describe the
weak-strong cluster model, followed by a summary of our results in
Sec.~\ref{sec:results}.  Concluding remarks are summarized in
Sec.~\ref{sec:conclusions}.  All the algorithms used in this
study are outlined in the \appendixname.

\section{Limited quantum speedup redefined}
\label{sec:qs}

Given the intrinsic differences between classical and quantum
heuristics, it is impossible to define a simple recipe to quantify
``quantum speedup.'' In Ref.~\cite{ronnow:14a}, the authors
discuss in detail the meaning of quantum speedup, defining different
``classes'' of speedup to better quantify any putative speedup of a
quantum device \cite{comment:analogy}.  More precisely, they classify
quantum heuristics in four different classes, ranging from the class
with the \emph{strongest} proof of quantum enhancement to the class with
the \emph{weakest} proof, as follows:\\

\noindent {\em Provable quantum speedupi.} It is rigorously proven that
no classical algorithm can scale better than a given quantum algorithm.
For example, the Grover algorithm (assuming an oracle) \cite{grover:97}
belongs to this class.\\

\noindent {\em Strong quantum speedup.} Originally defined in
Ref.~\cite{papageorgiou:13}, strong quantum speedup refers to
a comparison with the best classical algorithm, regardless if the
algorithm exists or not. Note, however, that the ``best classical
algorithm'' might not be known or there might be no consensus as to what
the best classical algorithm is. For example, the well-known Shor
quantum algorithm for the factorization of prime numbers \cite{shor:97}
belongs to this class.\\

\noindent {\em Potential quantum speedup.} Refers to speedup when
comparing to a specific classical algorithm or a set of classical
algorithms. In this case, any potential quantum speedup might be
short lived if a better classical algorithm is developed. An example is
the simulation of the time evolution of a quantum system, where the
propagation of the wave function on a quantum computer would be
exponentially faster than the direct integration of the Schr\"odinger
equation.\\

\noindent {\em Limited quantum speedup.} Speedup obtained by comparing
the algorithmic approach used in a quantum computer to the closer
classical counterpart. Usually, quantum Monte Carlo (QMC) is used for
the comparison with adiabatic quantum optimization
\cite{santoro:02,boixo:14,ronnow:14a}.\\

The introduction of the aforementioned categories has helped enormously
in ensuring that there are no misunderstandings when referring to
quantum speedup. Indeed, these general categories have the advantage
that they cover a broad class of quantum computing paradigms. However,
given that, at the moment, analog quantum annealing machines dominate
this field of research, it might be of importance to introduce 
definitions for quantum speedup tailored towards these machines.
Therefore, to be able to perform a fair assessment of speedup for
quantum annealing machines, we introduce the following definitions that
complement the notion of ``{\em limited quantum speedup}:''\\

\noindent {\em Limited sequential quantum speedup.} Speedup obtained
by comparing a quantum annealing algorithm or machine to any {\em
sequential} algorithm [e.g., simulated annealing (SA)
\cite{kirkpatrick:83}, or population annealing (PA) Monte Carlo
\cite{hukushima:03,machta:10,wang:15,wang:15e}] where a control
parameter is monotonously tuned until a certain threshold is reached
(e.g., the temperature in SA or the transverse field in QA). While
sequential methods might not necessarily be the best classical
optimization algorithm, they are the classical counterpart to quantum
annealing.\\

\noindent {\em Limited nontailored quantum speedup.} Speedup obtained
by comparing a quantum annealing algorithm or machine to the best-known
{\em generic} classical optimization algorithm that is not tailored to a
particular problem and does not exploit particular knowledge of the
problem to be optimized [e.g., isoenergetic cluster optimizers (ICM)
\cite{zhu:15b}, or the groups method \cite{zintchenko:15}].\\

\noindent {\em Limited tailored quantum speedup.} Speedup obtained by
comparing a quantum annealing algorithm or machine to the best-known
{\em tailored} classical optimization algorithm that explicitly exploits
the structure of the problem to be optimized and will perform in a
sub-optimal fashion (if work at all) on any other type of optimization
problem [examples are hybrid cluster moves (HCM) \cite{venturelli:15a}
or the Hamze--de-Freitas--Selby (HFS) algorithm
\cite{hamze:04,selby:14}].\\

Given the sequential nature of transverse-field quantum annealing, {\em
limited sequential quantum speedup} is naturally the fairest comparison
to classical counterparts. However, this might not be of much use if
classical sequential algorithms are slow compared to other classical
optimization methods. A comparison to tailored algorithms is slightly
unfair, because the structure of the problem is being exploited, i.e.,
the developer of the algorithm knows {\em a priori} how to design the
algorithm to outperform quantum annealing. We do emphasize that it might
be misleading to compare limited tailored quantum speedup to potential
quantum speedup because the classical algorithm is specifically designed
to outperform the quantum counterpart. However, comparing to nontailored
classical algorithms is similar to potential quantum speedup. The
classical approach is generic and widely applicable and makes no
assumptions about the studied problem. In addition, it should be the
currently fastest optimizer available \cite{comment:firas}.

Finally quantum annealing with a transverse field is the simplest
possible quantum-enhanced algorithm. Going beyond more complex driving
Hamiltonians (e.g., non-stoquastic \cite{matsuda:09}, different initial
states \cite{crosson:14}, the insertion of Hamiltonians during the
annealing \cite{perdomo:11}, or schedule randomization
\cite{comment:troyer}), one could easily imagine implementing far more
complex quantum algorithms that exploit the current advantages of
classical methods. For example, quantum cluster updates can be
implemented by suitably coupling two systems with the same target
Hamiltonian together, or quantum population annealing by running
multiple quantum chips in parallel and culling the least fit copies of
the target Hamiltonian. Once the field of quantum optimization has
reached this stage of development, the aforementioned defined categories
will require adjustments to take into account these advances.

\section{Weak-Strong cluster model}
\label{sec:wscm}

\begin{figure}
\includegraphics[width=\columnwidth]{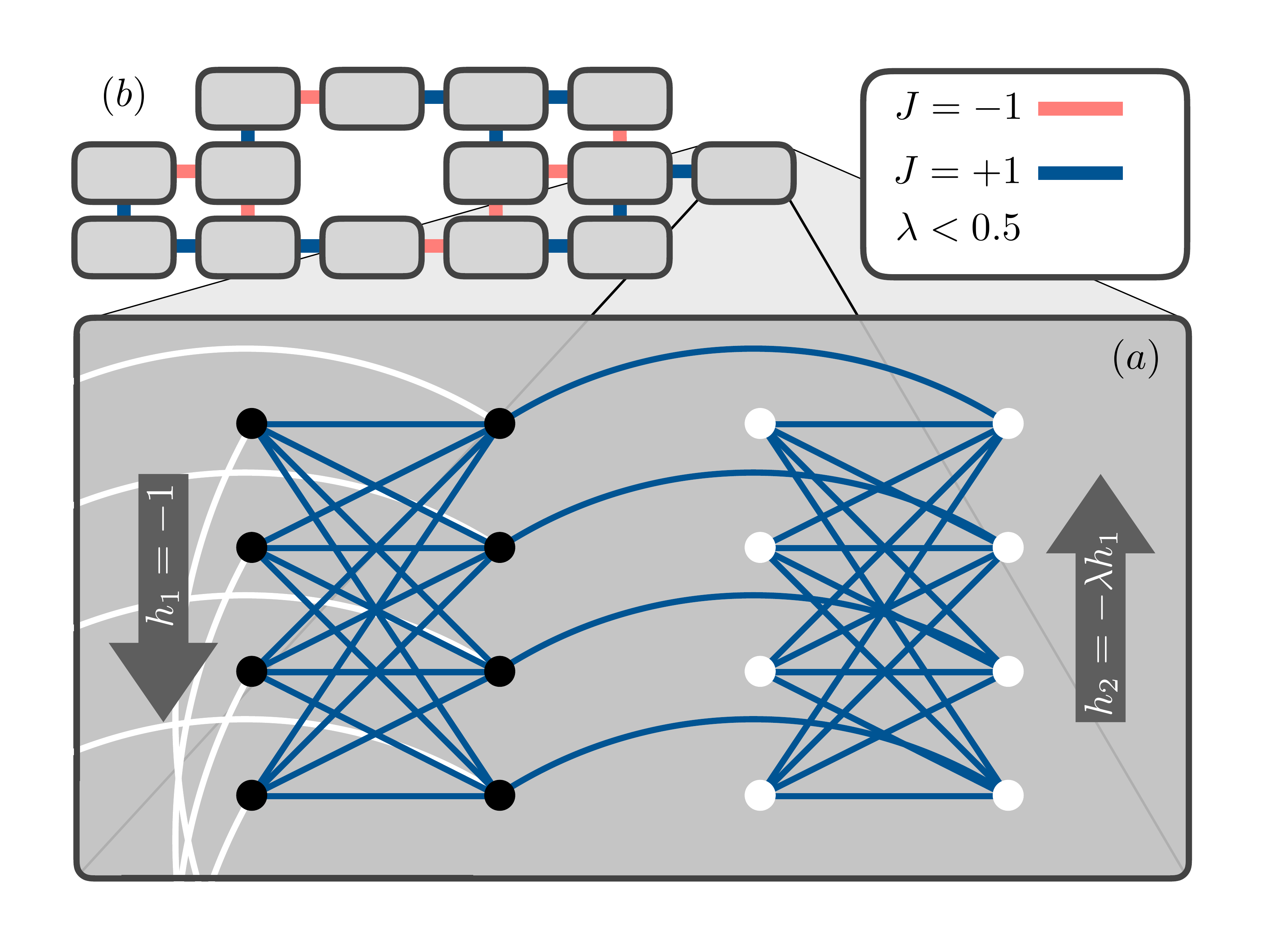}
\caption{\label{fig:weak-strong}
Sketch of the weak-strong clusters and networks.
(a) Structure of a weak-strong cluster. Two \kff cells of the chimera
lattice are connected ferromagnetically (blue lines, $J = 1$), as well
as all spins within each \kff cell. Black dots correspond to qubits in
the strong cluster with a biasing magnetic field $h_1 = -1$. The white
dots represent the weak cluster, where each site is coupled to a
weaker field $h_2 = -\lambda h_1$ with $\lambda = 0.44 < 0.5$ in the
opposite direction. The white lines represent the connections from the
strong cluster to neighboring strong clusters of a weak-strong pair. (b)
Weak-strong cluster network: each rectangle represents a weak-strong
cluster. The different weak-strong clusters are connected via a
spin-glass backbone where the interactions can take values $J_{\bar{x}\bar{x}^\prime} = \pm 1$. 
Here, red lines represent $J = -1$. Note that the connections between 
clusters only occur between the strong clusters.}
\end{figure}

The weak-strong cluster network is a tailored model designed to exploit
quantum tunneling in quantum optimizers and, therefore, to demonstrate
how finite-range tunneling can provide a computational advantage over
classical heuristics \cite{denchev:16}.  The model is composed by
highly-connected and ferromagnetically coupled clusters ($J = 1$)
(corresponding to the unit cells of the chimera graph \cite{bunyk:14})
that interact with each other (see Fig.~\ref{fig:weak-strong}).
These clusters can be divided in two classes: ``strong'' clusters, which
form the spin-glass bulk of the model, and ``weak'' clusters, which are
ferromagnetically coupled to strong clusters. To complete the model, an
external field is applied to all the spins of the system: a ``strong''
negative external field $h_1 = -1$ to those spins belonging to strong
clusters and a ``weak'' positive external field $h_2 = -\lambda h_1 =
0.44 < 1/2$ to those spins belonging to weak clusters. The ground state
of the system is therefore the configuration with all spins of both weak
and strong clusters pointing along the direction of the strong local
field. Individual weak-strong clusters are coupled by a spin-glass
backbone where the interactions between the clusters can take values 
$\{ \pm 1\}$. Note that the interactions between weak-strong clusters
only occur between sites in the strong cluster. See \figurename~3 of
\referencename~\cite{denchev:16} for the actual graphs simulated on the
\DW quantum annealer.  The peculiarity of the weak-strong cluster model
is that there exists a bifurcation point during both the classical and
quantum annealing where the system is forced to follow a ``wrong'' path
leading to a local minimum, namely the configuration with spins in weak
clusters pointing toward the weak external field. However, quantum
annealers, unlike classical annealers, can tunnel earlier to the
``correct'' path and, eventually, reach the true ground state of the
system.

The Hamiltonian describing the weak-strong cluster model is composed of
four main terms: the Hamiltonian describing the strong (weak) clusters
$\mathcal{H}_1$ ($\mathcal{H}_2$) and the Hamiltonians describing either
the interaction between strong clusters $\mathcal{H}_{1,1}$ or the
interaction between strong and weak clusters $\mathcal{H}_{1,2}$ (see
\referencename~\cite{denchev:16} for details). Each pair of weak-strong
cluster can be seen as a single functional cluster [i.e. a single gray
box in \figurename~\ref{fig:weak-strong}(b)], labeled by a
two-dimensional spatial position $\bar{x}$. Strong clusters belonging to
two different functional cluster are then linked with random couplings
($J_{\bar{x}\bar{x}^\prime} = \pm 1$) following a pre-determined backbone $\mathcal{B}$.  Note
that the weak clusters {\em only} couple to the strong cluster within a
given weak-strong cluster.  More precisely, for $\ell = \{1,\,2\}$ the
aforementioned Hamiltonians have the form
\begin{equation}
{\mathcal{H}}^{\bar{x}}_{\ell} = -J \sum_{i, j \in {\mathcal{V}_{\bar{x}}}} \sigma_{\ell,i}^z \sigma_{\ell,j}^z
- \sum_{i \in {\mathcal{V}}_{\bar{x}}} h_\ell \sigma_{\ell,i}^z
\label{eq:ham-intra}
\end{equation}
and
\begin{subequations}\label{eq:ham-inter}
\begin{align}
{\mathcal{H}}_{1,1}^{\bar{x},\bar{x}^\prime} &= - \sum_{j \in {{\mathcal{V}}}_{\bar{x},\bar{x}^\prime}} 
J_{\bar{x},\bar{x}^\prime}\sigma_{1,j}^z \sigma_{1,j}^z,\\
{\mathcal{H}}_{1,2}^{\bar{x}} &= - J \sum_{j \in \tilde{{\mathcal{V}}}_{\bar{x}}} \sigma_{1,j}^z \sigma_{2,j}^z,
\end{align}
\end{subequations}
where $\mathcal{V}_{\bar{x}}$ represents the eight vertices in one \kff
unit cell of the chimera graph for the functional cluster in the
position $\bar{x}$.  The set ${\mathcal{V}}_{\bar{x},\bar{x}^\prime}$
represents the vertices on the left-hand side which couple two adjacent
strong clusters while the set $\tilde{{\mathcal{V}}}_{\bar{x}}$
represents the vertices of the right-hand side of the strong and weak
clusters that are linked by a ferromagnetic interaction $J = 1$. Putting
together Eqs.~(\ref{eq:ham-intra})~and~(\ref{eq:ham-inter}), the
final problem Hamiltonian for the weak-strong cluster model assumes the
form:
\begin{equation}
  \mathcal{H} = \sum_{\bar{x}\in\mathcal{B}}\Big[
    \mathcal{H}^{\bar{x}}_{1} + 
    \mathcal{H}^{\bar{x}}_{2} + 
    \mathcal{H}^{\bar{x}}_{1,2}
  \Big] + 
  \sum_{(\bar{x},\bar{x}^\prime)\in\mathcal{B}} \mathcal{H}_{1,1}^{\bar{x},\bar{x}^\prime},
\end{equation}
with $(\bar{x},\bar{x}^\prime)$ indicating two functional clusters which
are adjacent in the given backbone $\mathcal{B}$.  Because of
imperfections in the \DW device, the embedding of the weak-strong
cluster network in the chimera topology is nontrivial. However, systems
of up to $n = 945$ qubits have been studied.

The main result of Ref.~\cite{denchev:16} is to show, either
experimentally (by using the \DW quantum optimizer) or numerically (by
using quantum Monte Carlo simulations), that quantum co-tunneling
effects play a fundamental role in adiabatic optimization.  Note that
quantum Monte Carlo is the closest classical algorithm to quantum
annealing on the DW2X.  The results of Ref.~\cite{denchev:16}
on the \DW chip are approximately $10^8$ times faster than simulated
annealing \cite{kirkpatrick:83} and considerably faster than quantum
Monte Carlo despite both the \DW quantum annealer and quantum Monte
Carlo having a similar scaling (similar slope of the curves in
Fig.~4 of Ref.~\cite{denchev:16} for quantum Monte
Carlo and the DW2X).  While this, indeed, represents the first solid
evidence that the \DW machine might have capabilities that classical
optimization approaches do not possess, it is important to perform a
comprehensive comparison to a wide variety of state-of-the-art
optimization methods.  Within the categories defined in
Sec.~\ref{sec:qs}, the results of
Ref.~\cite{denchev:16} for the \DW clearly outperform any
sequential optimization methods, however, fall short of outperforming
tailored and nontailored optimization methods. We feel, however, that
knowingly exploiting the structure of a problem does not amount to a
fair comparison. However, our results shown below clearly suggest that
generic optimization methods still outperform the DW2X. One might thus
question the importance of the results of
\referencename~\cite{denchev:16}. We emphasize that this is the first
study that undoubtedly shows that the \DW machine has finite-range
tunneling and gives clear hints towards the class of problems where
analog quantum annealing machines might excel.

In addition to showing here that a variety of either ``tailored'' to the
weak-strong cluster structure or more ``generic'' classical heuristics
can achieve similar performances of the \DW chip, we also study the
energy landscape of the weak-strong cluster networks. The latter
provides valuable insights about the limitations of finite-range
tunneling for this class of problems. Our analysis suggests that the
scaling advantage of finite-range cotunneling over sequential algorithms
could be lost for instances with problem sizes beyond the ones
considered  in Ref.~\cite{denchev:16}.

In the next section we further discuss the performance of \DW compared
to tailored and nontailored classical heuristics in detail.

\section{Results}
\label{sec:results}

In this section, we present our main results. In the first part, we
compare the performance of the \DW device against general (nontailored)
and tailored classical algorithms. The description of the used
algorithms is in the \appendixname. In the second part, we analyze in
depth the scaling behavior of the \DW device by varying the number of
used qubits. The aim is to better understand the role of nonoptimal
annealing times for a noisy analog device to the asymptotic scaling of
the computational time. Finally, we study the energy landscape, as
proposed in Ref.~\cite{katzgraber:15}, and show that for
increasing problem size the spin-glass backbone of the weak-strong
cluster network dominates and the advantages of finite-range tunneling
diminish for increasing system sizes.

\subsection{Analysis of the computational scaling}
\label{subsec:comp_scaling}

In order to compare heuristics which are fundamentally different from
each other, it is necessary to define a metric which is not only
\emph{fair}, but that gives a \emph{quantitative} measure of the
speedup.  In this work, and to compare on equal footing with the results
in \referencename~\cite{denchev:16}, we follow the
\emph{time-to-solution} metric introduced in
Refs.~\cite{boixo:14,ronnow:14a}.  This metric is defined as the time to
find the ground state with $99\%$ of confidence after a given number of
repeated runs, namely,
\begin{equation}\label{eq:tts}
	T_\text{tts} = T_\text{ann} R = T_\text{ann}\frac{\log_{10}(1 - 0.99)}{\log_{10}[1 - p_\text{succ}(T_\text{ann})]},
\end{equation}
where $T_\text{ann}$ is the annealing (running) time of the quantum
(classical) heuristic and $R$ is the number of repetitions needed to
reach a confidence $s$. For the current generation of the DW2X, the
total annealing time $T_\text{ann}$ cannot be arbitrarily small. For the
experiments described here, $ T_\text{ann}$  was set to the minimum time
allowed in the device ($20\,\mu$s). In the next section, we better
describe the consequences imposed by this limitation to correctly
extrapolate the asymptotic limit of the computational time.

In general, we are interested in the asymptotic limit of the
computational time $T_\text{tts}$ to understand what would be
the true scaling for large systems. For the weak-strong cluster network, 
it is expected that $T_\text{tts}$ grows exponentially with the system
size (up to a polynomial correction) as:
\begin{equation}\label{eq:tts_limit}
	T_\text{tts} \approx \text{poly}(n)\, 10^{a + b\,\sqrt{n}} = 10^{a + b\,\sqrt{n} + c\log_{10}\left(\sqrt{n}\right)},
\end{equation}
with $n^{c/2}$ the dominant term of the polynomial prefactor
$\text{poly}(n)$. Observe that, for the scaling in
Eq.~(\ref{eq:tts_limit}), we choose $\sqrt{n}$ rather than
$n$. This choice has been made for two main reasons.  On one hand, it is
well known that optimization problems on xhimera Hamiltonians have a
computational scaling that it is well approximated by a stretched
exponential \cite{boixo:14,ronnow:14}. On the other hand, the graph
underlying the chimera Hamiltonian is almost planar with a treewidth
equal to $\sqrt{n}$ (as a two-dimensional lattice) rather than $n$ (as a
fully connected graph) \cite{selby:14}. Hence, it is expected that
typical collective excitations involve a number of qubits of the order
of $\sqrt{n}$.  Among all the parameters in
Eq.~(\ref{eq:tts_limit}), the most important parameter is $b$
because it represents the dominant term in the limit of large systems.
In order to determine the values of parameters $a$, $b$, and $c$ in
Eq.~(\ref{eq:tts_limit}), it is possible to either use a
linear fit of the form
\begin{equation}
f(x) = a + b\,\sqrt{n} ,
\end{equation}
i.e. it is assumed that the term $c$ is
negligible, or a log-corrected fit  of the form
\begin{equation}
f(x) = a + b\,\sqrt{n} + c\log_{10}\left(\sqrt{n}\right) .
\end{equation}
The advantage of a linear fit is that less parameters have to be
determined.  However, it is more affected by finite-size effects.  The
log-corrected regression, on the contrary, takes into account eventual
finite-size effects but the regression could display a ``non-physical''
scaling behavior for small system sizes where the fit increased for $n
\to 0$ (see, for instance, the top-left panel of
Fig.~\ref{fig:fit_50}).

\begin{figure}
\includegraphics[width=0.48\textwidth]{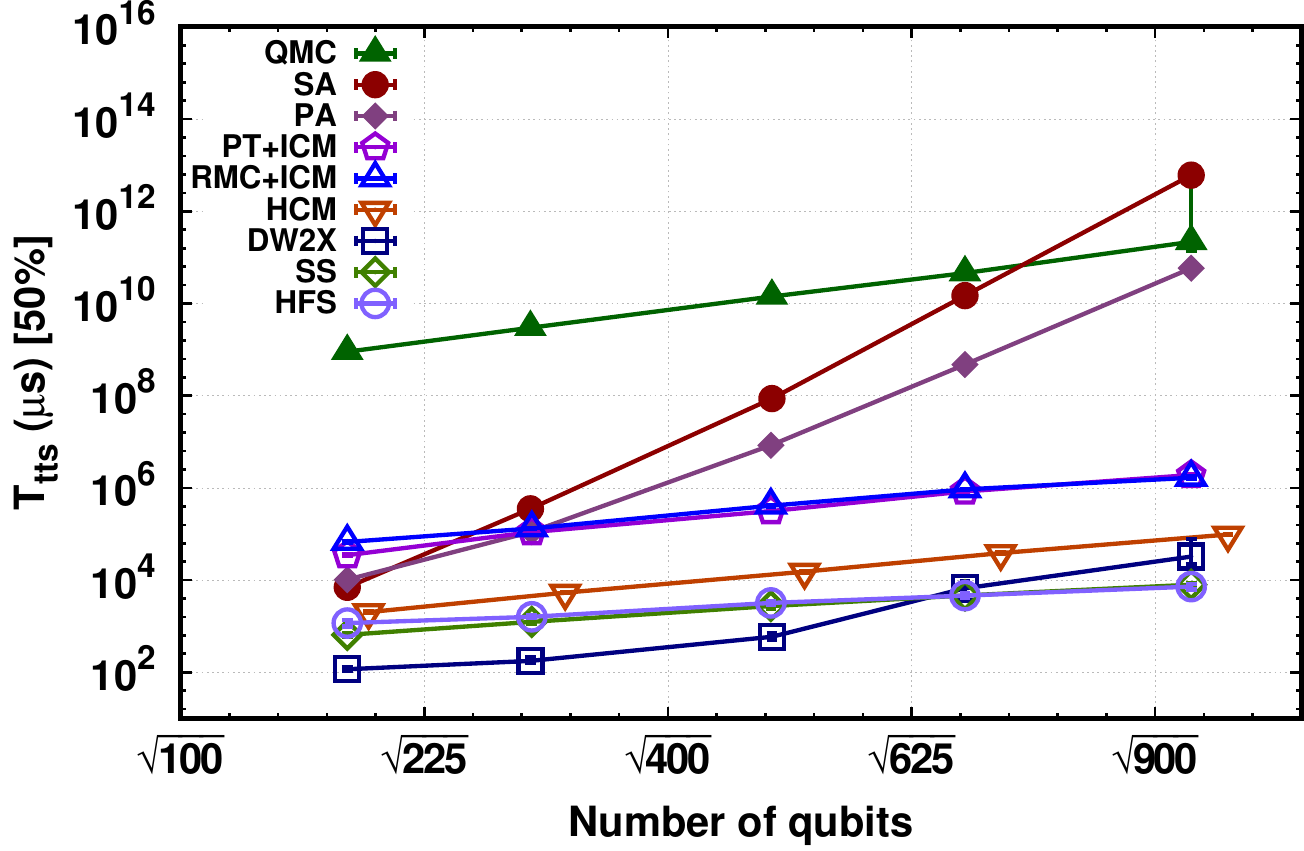}
\includegraphics[width=0.48\textwidth]{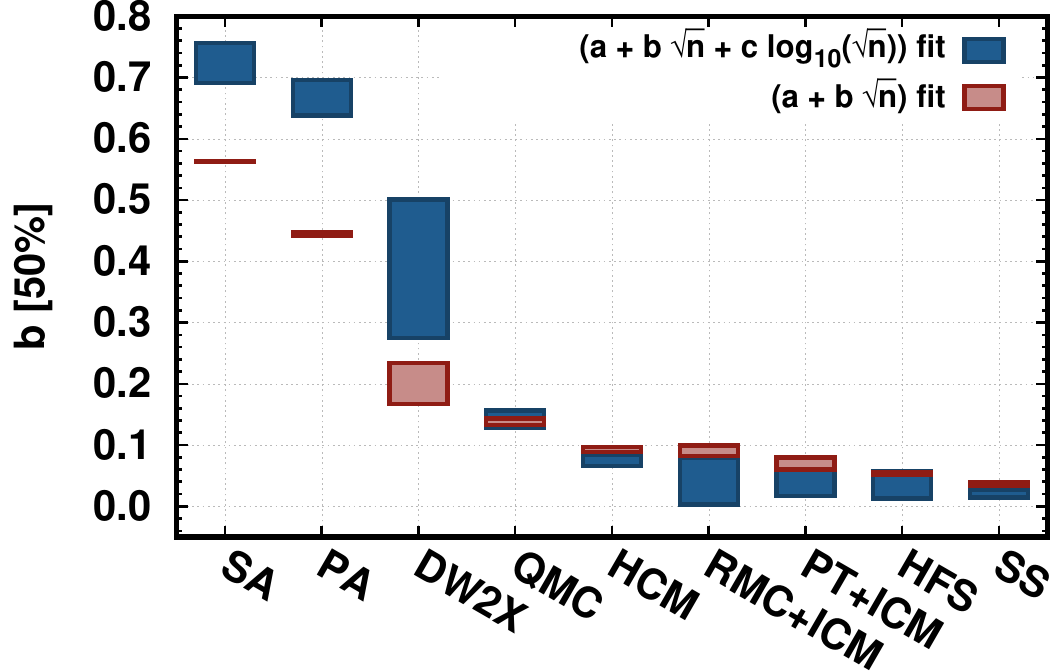}
\caption{\label{fig:scaling_50}
Top panel: computational scaling (for 99\% success) for different
classical algorithms compared with the experimental results using the
\DW chip \cite{denchev:16}. As one can see, both general classical
algorithms [isoenergetic cluster moves (ICM) either using parallel
tempering (PT) or replica Monte Carlo (RMC)] and tailored classical
algorithms for the weak-strong cluster model [hybrid cluster moves
(HCM), super-spin approximation (SS), Hamze--de-Freitas--Selby (HFS)] have
a comparable scaling with the quantum inspired classical algorithm
[quantum Monte Carlo (QMC)] and the \DW device.  \cite{comment:sap}.
Moreover, for the classical tailored algorithms, the overall scaling
prefactor is also comparable with the \DW device.  For HCM, random
instances with \emph{no} broken qubits have been used.  Bottom panel:
analysis of the scaling factors by using either linear regression, or a
log-corrected regression for $\log_{10} T_\text{ann}$. In the figure,
bars represent the confidence intervals. For the scaling analysis, we
used a stretched exponential that fits better the numerical data (see
Appendix \ref{app:ana}).  Interestingly, the general-purposes classical
algorithm ICM, together with the chimera-optimized classical algorithm
(HFS) and the cluster-optimized algorithms (HCM and SS) have the best
scaling.  (QMC and SA data taken from Ref.~\cite{denchev:16}.) All the
simulations (excluding HCM) have been run on the same instances used in
\cite{denchev:16}.
}
\end{figure}

In Fig.~\ref{fig:scaling_50}, we report the computational scaling
of the various classical and quantum heuristics considered in this paper
(top panel), as well as the asymptotic parameter $b$ (bottom panel).
The results show that sequential quantum approaches (\DW and QMC)
clearly outperform classical sequential algorithms [simulated annealing
(SA) and population annealing(PA)], having a smaller asymptotic scaling
exponent $b$.  Nevertheless, both tailored [hybrid cluster method (HCM),
Hamze--de-Freitas--Selby (HFS) and the super-spin approximation (SS)] and
nontailored classical algorithms [isoenergetic cluster moves (ICM)
combined with either parallel tempering (PT+ICM) or replica Monte Carlo
(RMC+ICM)] have a better performance.

We emphasize that these results are specific to the \DW quantum annealer
and its underlying chimera topology. Certain nontailored algorithms
might not perform as well on different topologies or other problem
classes. For example, the general classical ICM algorithm in its native
implementation \cite{zhu:15b} would not be as efficient for
highly connected graphs. Therefore, future quantum annealing machines
with denser connectivities might again outperform the current classical
state of the art, at which point, hopefully, more efficient classical
methods will be developed.

\subsection{Non-optimal annealing time and ``double scaling''}
\label{subsec:two-level-system}

In the previous section, we analyzed the performance of the various
classical and quantum heuristics by looking at the computational
scaling. More precisely, we are interested in the asymptotic behavior of
the time to solution [see Eq.~(\ref{eq:tts})] that it is
expected to be exponential in the limit of large systems:
\begin{equation}\label{eq:tts_limit2}
	T_\text{tts} \approx e^{-b \sqrt{n}},
\end{equation}
where $b$ is the asymptotic scaling exponent. However, how large should
the system be in order to extrapolate the asymptotic scaling $b$? Many
factors such as the annealing schedule \cite{roland:02, mandra:15}, as
well as the intrinsic noise of the system \cite{mandra:15,
shenvi:03,roland:05, amin:08}, can affect the scaling behavior of the
computational time. To address the above question, we show in this
section that the use of a non-optimal annealing schedule can lead
to a ``double-scaling'' effect where the true asymptotic scaling is
hidden by a fictitious (but more favorable) scaling.

It is well known that the computational scaling of a quantum annealer
represents only an upper bound of the true scaling if a non-optimal
schedule is used~\cite{ronnow:14,hen:15a}. For instance, consider the
case of a fixed schedule but with a very large annealing time. In this
case, the computational scaling would be a flat curve because the
probability of success would be one for almost all system sizes
available for examination. Therefore, very large systems are required to
extrapolate to the correct asymptotic scaling.

The \DW quantum annealer has a fixed schedule and, as previously
mentioned, a minimum annealing time of $20\,\mu$s.  Furthermore, the \DW
chip is affected by an unavoidable intrinsic noise
\cite{venturelli:15a,katzgraber:15,zhu:16,perdomo:15} that can alter the
computational scaling.

\begin{figure}
\includegraphics[width=0.48\textwidth]{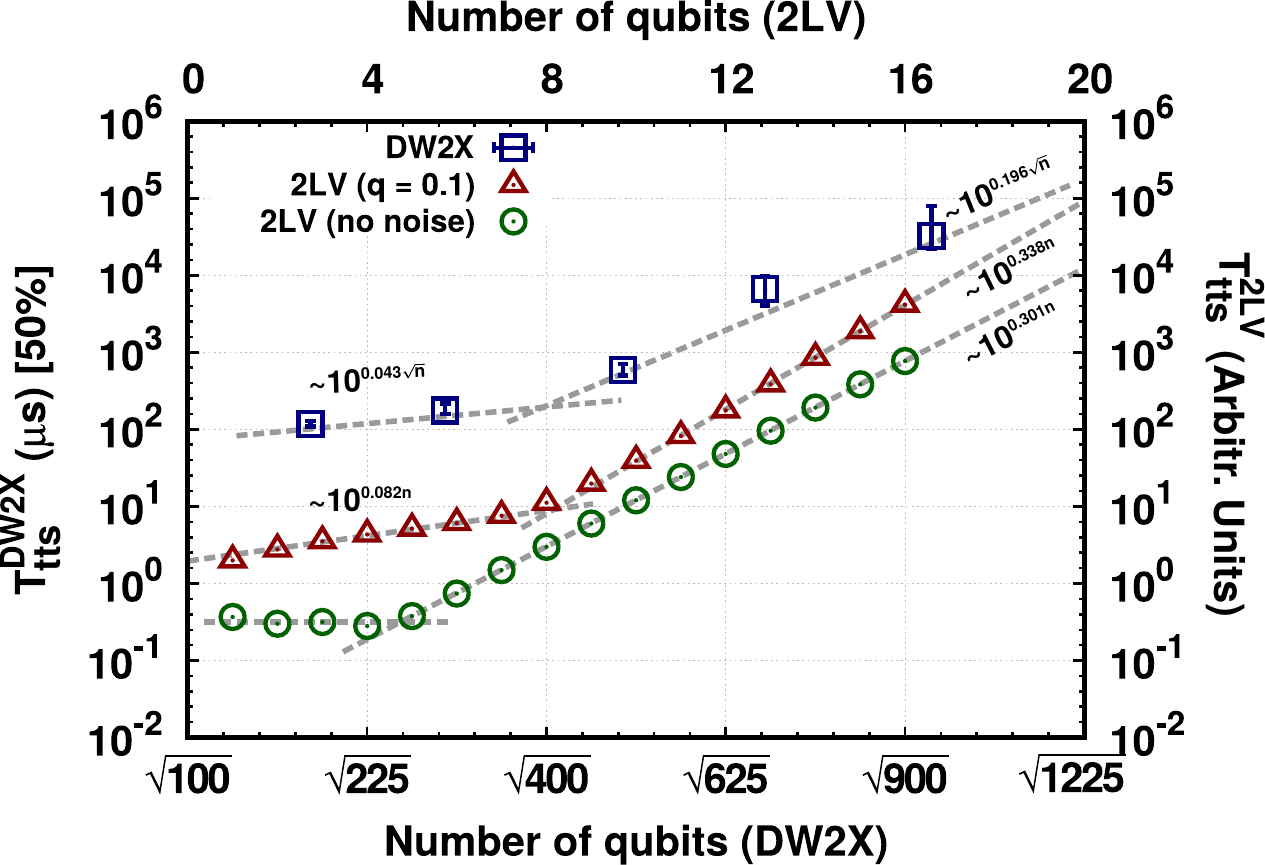}
\caption{\label{fig:2LV-scaling}
Double-scaling effect produced by the combination of a noisy system
and a non-optimal annealing time.
We display data for the \DW device (as
in \cite{denchev:16}) compared to the annealing of a noisy two-energy level
model (2LV) with a non-optimal (linear) annealing schedule and a fixed total
annealing time ($T_\text{ann} = 500$). The numerical study shows that, for small systems, the
scaling is mainly dominated by the noise while, for large systems, the
scaling is mainly dominated by the asymptotic behavior.}
\end{figure}

To better understand the scaling behavior of the \DW for the
weak-strong cluster model, we compare its scaling with the scaling
behavior of a noisy two-energy level model with a fixed (linear) schedule and
a non-optimal annealing time. More precisely, we use the following
Hamiltonian \cite{roland:02}:
\begin{equation}\label{eq:ham_2LV}
	\mathcal{H}_{\rm 2LV}(t) = -(1 - t/T_\text{ann})\ketbra{\psi}{\psi} - t/T_\text{ann}\ketbra{\omega}{\omega},
\end{equation}
where $T_\text{ann}$ is the total annealing time, and $\ket{\psi}$ and
$\ket{\omega}$ are the equal superposition of all the states and the
target states one wants to find, respectively. The system in
Eq.~(\ref{eq:ham_2LV}) can be reduced to an effective $2\times
2$ matrix and, then, it can be exactly solved \cite{roland:02,mandra:15}.
To simulate the presence of local noise, we assume that each spin has a
probability $q$ to be oriented in the wrong direction after the
annealing of the system \cite{mandra:15}. Therefore, the effective noisy
Hamiltonian has a probability equal to $(1-q)^n$ that its ground state
$\omega^\prime$ is effectively the desired target state $\omega$.
Assuming that the level of noise is small enough compared to the
probability of success $p_\text{succ}(n,\,T_\text{ann})$ of the perfect
annealer (namely, when $T_\text{ann}$ is much larger than the optimal
annealing time), the probability of success of the noisy two-energy
level Hamiltonian can be written as:
\begin{equation}\label{eq:p_succ_noisy}
	p^\prime_\text{succ}(n,T_\text{ann},\,\,q) = (1 - q)^n p_\text{succ}(n,\,T_\text{ann}).
\end{equation}
Figure \ref{fig:2LV-scaling} shows the comparison between the
computational scaling $T_\text{tts}$ for the \DW chip \cite{denchev:16}
and the two-energy level model described above (for the numerical details,
see Appendix~\ref{sec:2LV}). For the latter, the
computational scaling is expressed in arbitrary units in order to
ease the comparison. As expected, the ideal two-energy level model without
noise (2LV, $q = 0$) has a plateau for small systems and, only for large
systems, the computational time shows the asymptotic scaling. When the
noise is added to the two-energy level model (2LV, $q = 0.1$) a ``double
scaling'' phenomenon appears: for small systems, the scaling is
dominated by the noise while, for large systems, the scaling is
dominated by the asymptotic scaling. Interestingly, the same phenomenon
can be clearly observed for the \DW scaling, indicating that the total
annealing time of $20\,\mu s$ is non-optimal for systems up to
$\sqrt{400}$ spins.

\subsection{Analysis of the energy landscape}
\label{subsec:landscape}

An important ingredient in assessing the value of weak-strong cluster
problems to detect quantum speedup is to study in detail the dominant
characteristics of the energy landscape.  In
Refs.~\cite{yucesoy:13,katzgraber:15} it was shown that the structure of
the overlap distribution of spin glasses \cite{binder:86,stein:13}
mirrors salient features in the energy landscape. Because there is no
spatial order in spin glasses, ``order'' is measured by comparing two
copies of the system with the same disorder (i.e., the same set of
interactions between the qubits and the same magnetic fields), but
simulated with independent Markov chains (i.e., each copy starts from a
different random initial condition). The spin overlap is defined as
\begin{equation}
q = \frac{1}{n}\sum_{j = 1}^n \sigma_j^{z,\alpha} \sigma_j^{z,\beta} ,
\label{eq:q}
\end{equation}
where the sum is over all sites $n$ on the network and $\alpha$ and
$\beta$ represent the two copies of the system. For a given set of
disorder $J_{\bar{x}\bar{x}^\prime}$, the overlap
distribution $P(q)$ will have a unique structure at low, but finite
temperatures $T \ll J$, $T > 0$. Generally speaking, the number of peaks
roughly mirrors the number of dominant valleys in the (free-) energy
landscape \cite{stein:13}. The distance between peaks, as well as their
width, can be associated with the Hamming distance between dominant
valleys and their width, respectively.  As shown in
\referencename~\cite{yucesoy:13}, the more structure the distribution
has, the larger the typical computational complexity is. Furthermore, as
shown in \referencename~\cite{melchert:16}, when the distribution only
has one dominant peak, there is either one dominant valley in the energy
landscape or a set of strongly-overlapping valleys separated by thin but
tall barriers. In the latter case, the barriers are so thin that the
features in $P(q)$ overlap strongly, i.e., the distribution cannot
differentiate the different valleys.  However, if there are multiple
well-defined features, dominant valleys in the energy landscape are
separated by thick barriers.

\begin{figure}
\includegraphics[width=0.48\textwidth]{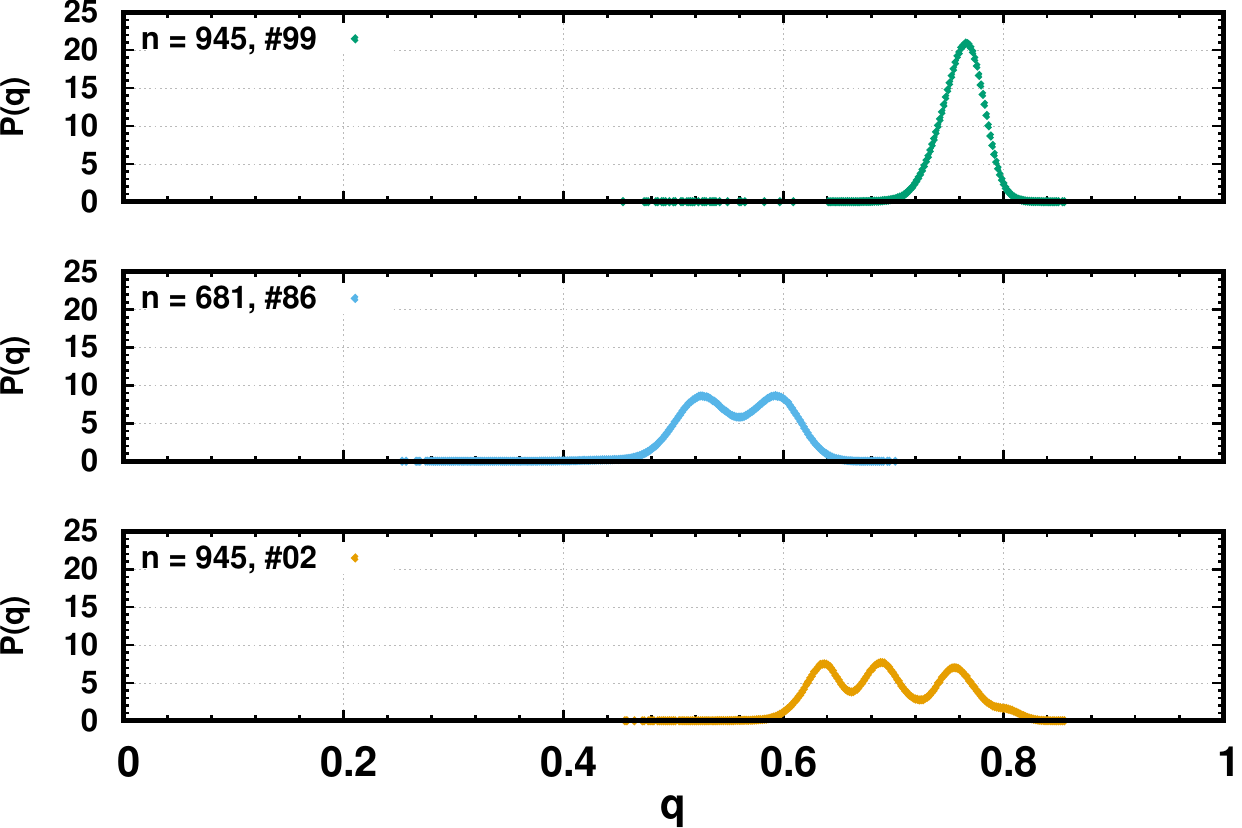}
\caption{\label{fig:pq}
Three representative overlap distributions $P(q)$ for different problem
sizes $n$.  The $y$-axes are in arbitrary units and rescaled such that
$\int_0^1 P(q) = 1$.  While some instances have either one dominant
narrow valley or valleys with thin barriers that allow for finite-range
tunneling (top), others have multiple structures (middle and bottom)
suggesting that the valleys are separated by barriers that might be too
wide for any finite-range tunneling to be beneficial during the
optimization.}
\end{figure}

Using parallel tempering Monte Carlo at low temperatures
\cite{katzgraber:15}, we have computed the overlap distribution for the
different weak-strong cluster networks. Because of the added fields,
there is no spin-reversal symmetry and the distributions only show peaks
for $q > 0$.  We find two characteristic shapes shown in
\figurename~\ref{fig:pq}: either the problems have a single dominant narrow
peak (compared to random spin-glass problems \cite{katzgraber:15}), or
multiple well-separated peaks. While the latter have energy
barriers that are too thick for any finite-range quantum tunneling to be
effective, the former potentially have thin enough barriers to allow for
finite-range tunneling in the DW2X. Therefore, only problems that have
single narrow peaks might benefit from any finite-range tunneling. With better
statistics, it would be instructive to study the scaling of both problem
classes separately with QMC and SA for systems considerably larger than
the DW2X.

\begin{figure}
\includegraphics[width=0.48\textwidth]{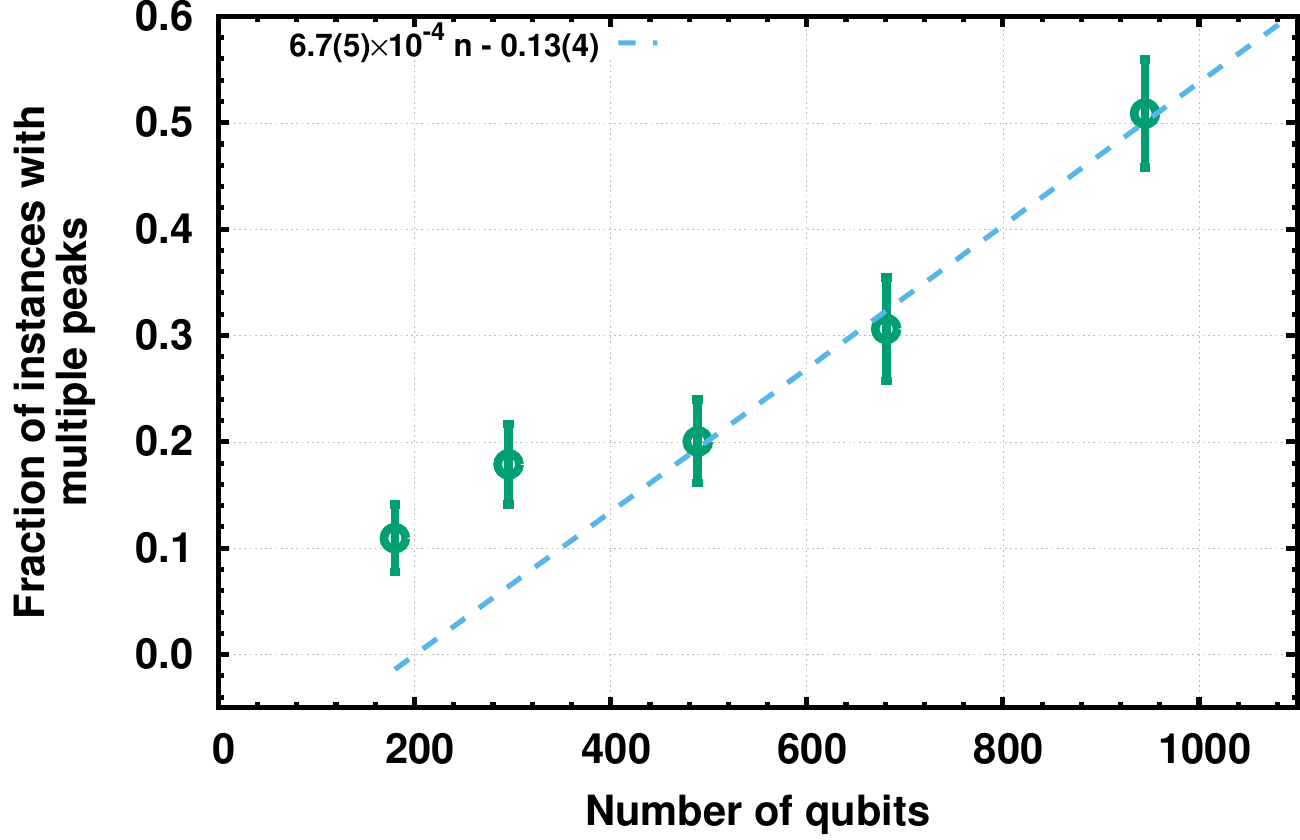}
\caption{\label{fig:pqratio}
Ratio of single peak to multi-peak overlap
distributions as a function of the number of qubits $n$.  For increasing
system size, multi-peak instances with barriers too wide for
finite-range tunneling increase due to the influence of the spin-glass
backbone. Note that already for the largest system size studied,
multi-valley instances dominate.}
\end{figure}

\figurename~\ref{fig:pqratio} shows the fraction of problems with multiple
peaks against problems with single peaks. The fraction of multi-peak
instances (problems with wide barriers in the energy landscape) grows
considerably with the problem size $n$, i.e., for large systems the
spin-glass backbone dominates and thus, asymptotically, finite-range
tunneling becomes inefficient on the DW2X. Loosely extrapolating the data
in Fig.~\ref{fig:pqratio}, we estimate that this class of problem
might show a change in scaling already for the next D-Wave chip
generation of approximately $2000$ qubits.

\section{Conclusions}
\label{sec:conclusions}

In this work, we study in detail and complement recent results by Google
Inc.~\cite{denchev:16} on the \DW quantum annealer.  Their results show
for the first time that a quantum annealing machine can outperform
conventional computing technologies for a particular class of problems.
However, to enable a more detailed comparison, we first expand the
notion of ``{\em limited quantum speedup}'' introduced in
Ref.~\cite{ronnow:14}.  In particular, to perform a fair
assessment of the results of Ref.~\cite{denchev:16}, we
introduce the notion of ``{\em limited sequential quantum speedup}''
which refers to a speedup over the best-known sequential algorithms, as
well as``{\em tailored and nontailored quantum speedup.}'' The latter
categories encompass numerical approaches that are not sequential and
either exploit the (known) structure of the optimization problem to be
solved or are generic. A strong yet fair indication for limited quantum
speedup would be to outperform the best-known generic algorithm. In the
case of the \DW when optimizing the weak-strong cluster networks, our
results show that while the \DW (as well as quantum Monte Carlo) has a
better scaling compared to sequential methods, tailored (as well as
nontailored) algorithms show a better asymptotic scaling.

Furthermore, as part of the study, we show that the role of the noise is
not marginal in the extrapolation of the asymptotic computational
scaling for large system sizes. More precisely, we explain the sudden
change of scaling of the computational time of the \DW device (and the
consequent effect of a ``double scaling'') by comparing the quantum
annealer with a noisy two-energy level model with a non-optimal
annealing schedule. In both cases, the true asymptotic scaling is hidden
by an initial (and more favorable) scaling, that later turns to the true
asymptotic scaling.

Finally, we study the dominant features in the energy landscape of the
weak-strong cluster network problems. Our results suggest that the
spin-glass backbone might dominate the scaling already for systems with
twice as many qubits as the current-generation \DW machine. As such, the
favorable speedup currently found both on quantum Monte Carlo
simulations as well as the \DW device might asymptotically approach
towards the scaling of the other sequential methods.

While one might see the results of \referencename~\cite{denchev:16} post
a detailed analysis presented in this paper as discouraging, we
emphasize that this is a careful study that has shown
strong results in favor of quantum annealing approaches both on analog
quantum annealing machines, as well as quantum simulations. Although
there is a clear evidence that random problems (e.g., spin glasses
\cite{ronnow:14}) might not be well suited for quantum annealing to
excel, tailored problems \cite{katzgraber:15} are of clear importance in
the quest of quantum speedup. Determining the application domain where
quantum annealing machines will surpass the capabilities of current
silicon-based technologies is of paramount importance across
disciplines, and the work by the Google Inc.~team has given the first
strong indications in which directions to search.

\begin{acknowledgments}

We thank the Google Quantum A.I.~Lab members for sharing their QMC and
SA data, multiple discussions, as well as making the weak-strong cluster
instances available to us. We also thank A.~Aspuru-Guzik, F.~Hamze,
A.~J.~Ochoa, and Eleanor G. Rieffel for many fruitful discussions, as
well as H.~Munoz-Bauza for help with the graphics.  H.~G.~K.~and
W.~W.~acknowledge support from the NSF (Grant No.~DMR-1151387).
H.~G.~K.~thanks Daniel Humm, Marco Pierre White, Thomas Keller, Heston
Blumenthal and Paul Bocuse for inspiration during the initial stages of
the manuscript. S.~M. was supported by NASA (Sponsor Award Number:
NNX14AF62G). We thank the Texas Advanced Computing Center (TACC) at The
University of Texas at Austin for providing HPC resources (Stampede
Cluster) and Texas A\&M University for access to their Ada, and Lonestar
clusters. This research is based upon work supported in part by the
Office of the Director of National Intelligence (ODNI), Intelligence
Advanced Research Projects Activity (IARPA), via MIT Lincoln Laboratory
Air Force Contract No.~FA8721-05-C-0002.  The views and conclusions
contained herein are those of the authors and should not be interpreted
as necessarily representing the official policies or endorsements,
either expressed or implied, of ODNI, IARPA, or the U.S.~Government.
The U.S.~Government is authorized to reproduce and distribute reprints
for Governmental purpose notwithstanding any copyright annotation
thereon.

\end{acknowledgments}

\appendix

\section{Analysis of the computational scaling}
\label{app:ana}

\begin{figure}[t!]
\includegraphics[width=0.45\textwidth]{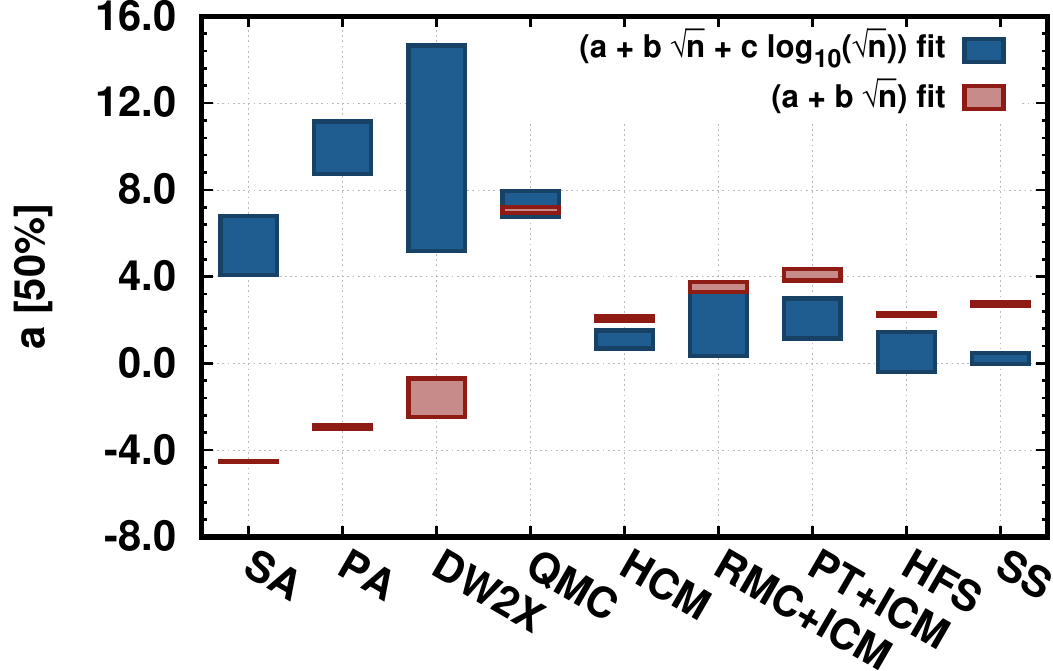}
\includegraphics[width=0.45\textwidth]{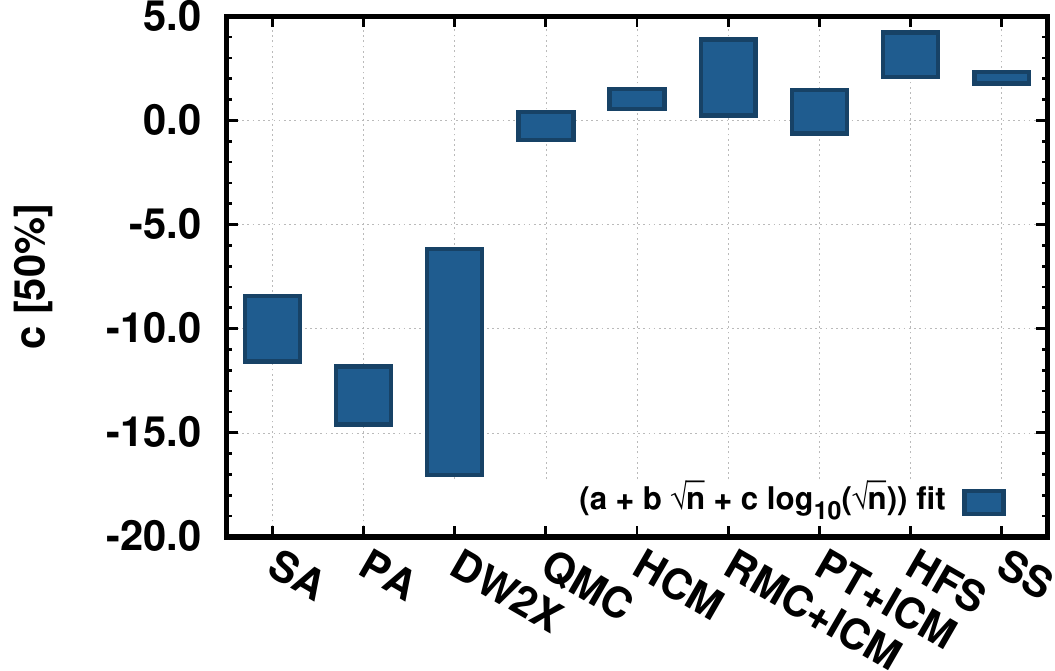}
\caption{\label{fig:factors_ab_50}{Values for the
remaining fit parameters $a$ and $c$ for either the log-corrected fit $a + b
\sqrt{n} + c \log_{10}(\sqrt{n})$ or the linear fit $a + b \sqrt{n}$}.
In the panels, bars represent the confidence interval.}
\end{figure}

\begin{figure*}[t!]
\includegraphics[width=0.3\textwidth]{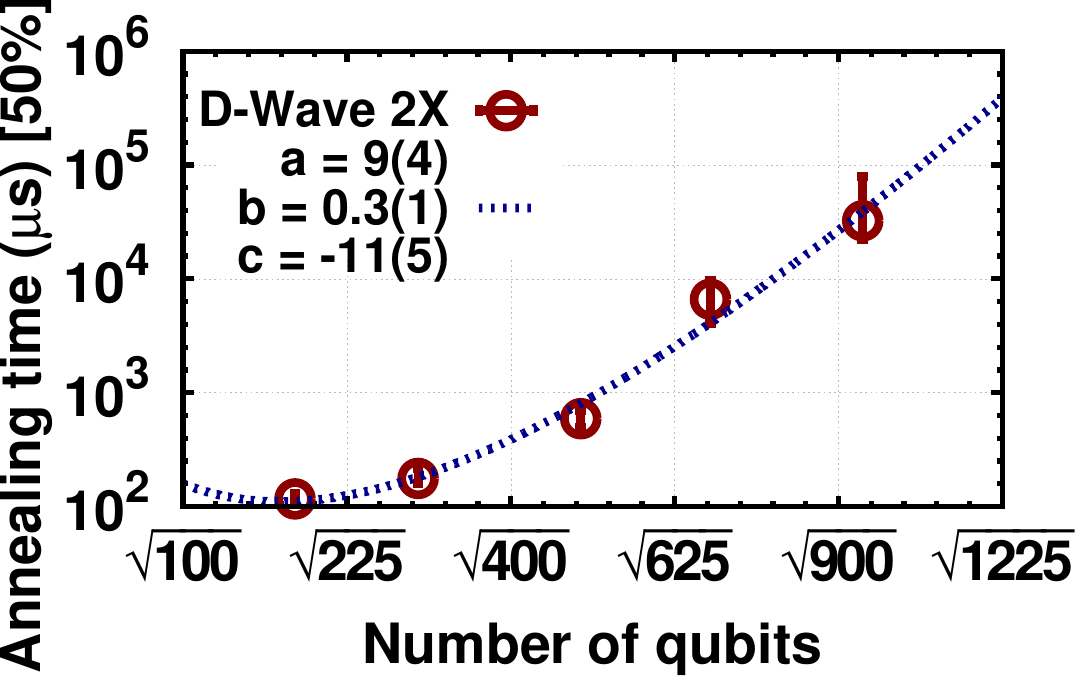}
\includegraphics[width=0.3\textwidth]{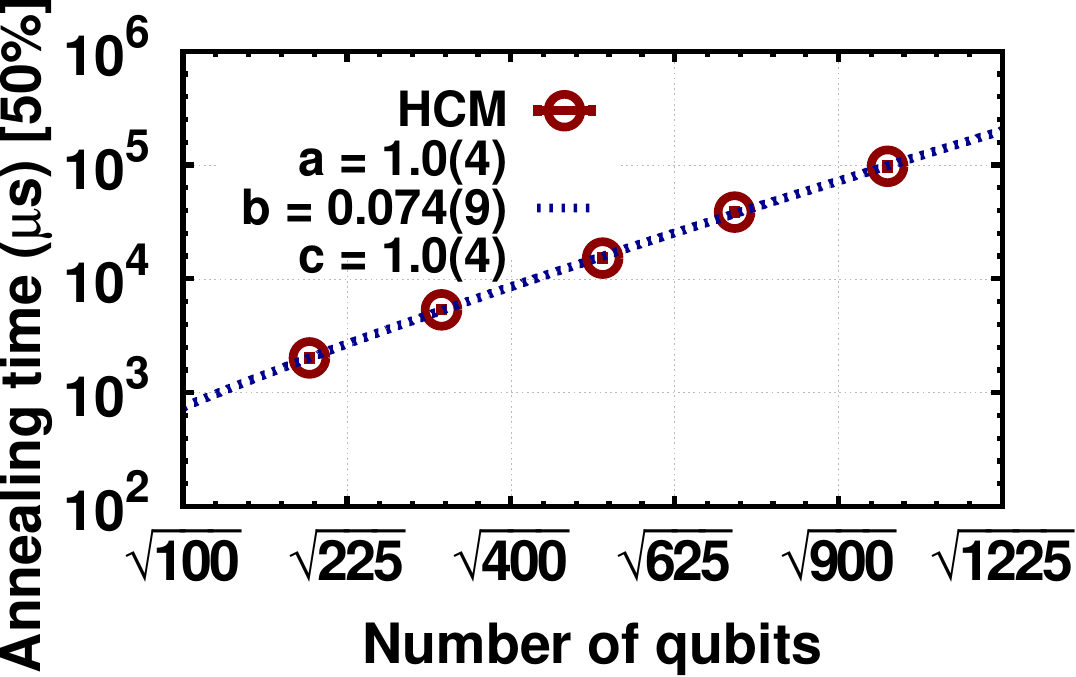}
\includegraphics[width=0.3\textwidth]{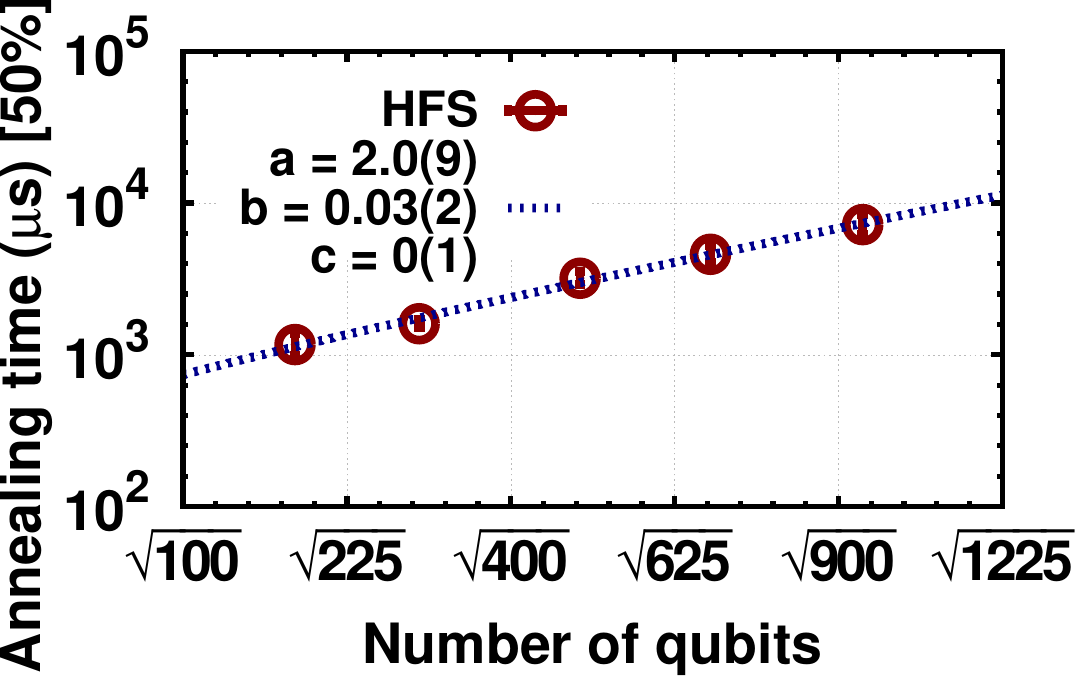}\\
\includegraphics[width=0.3\textwidth]{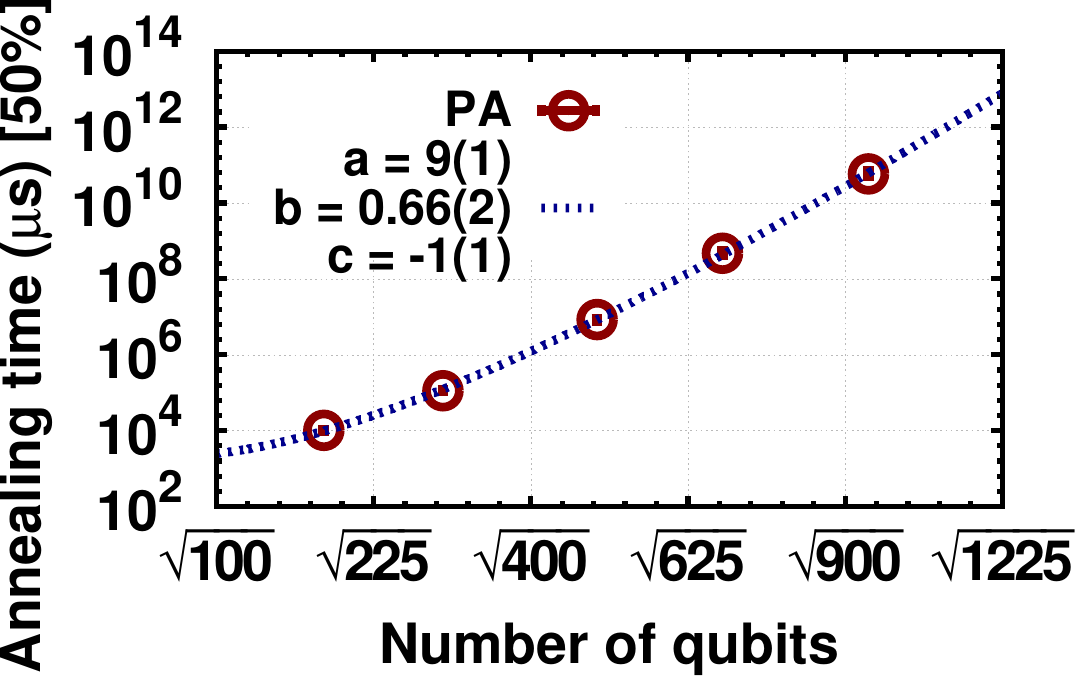}
\includegraphics[width=0.3\textwidth]{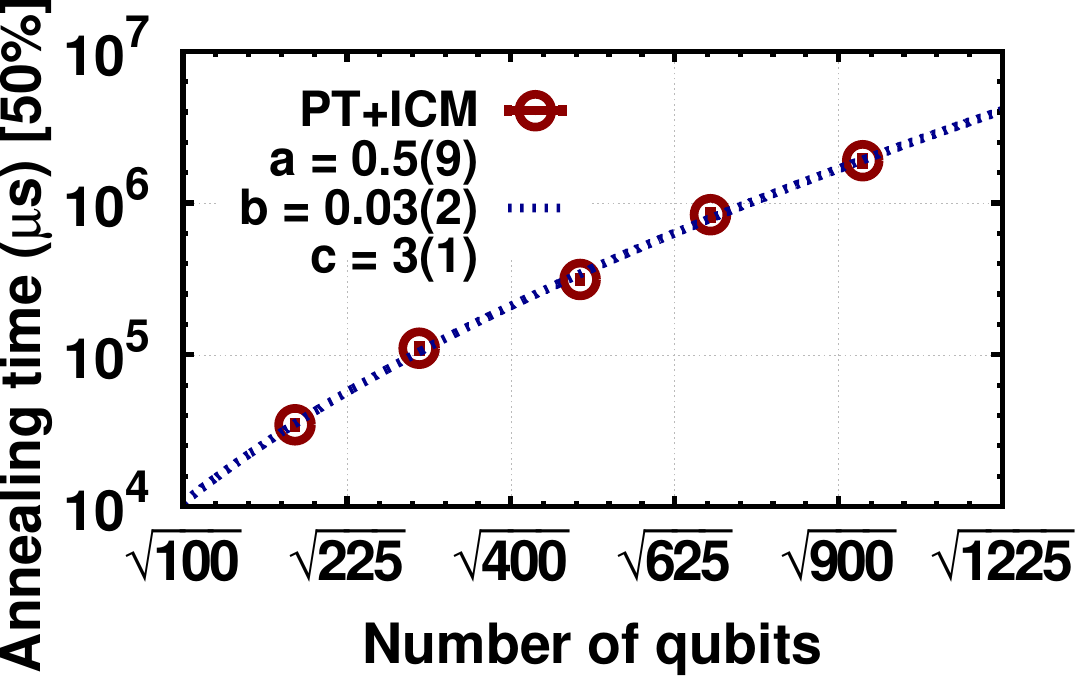}
\includegraphics[width=0.3\textwidth]{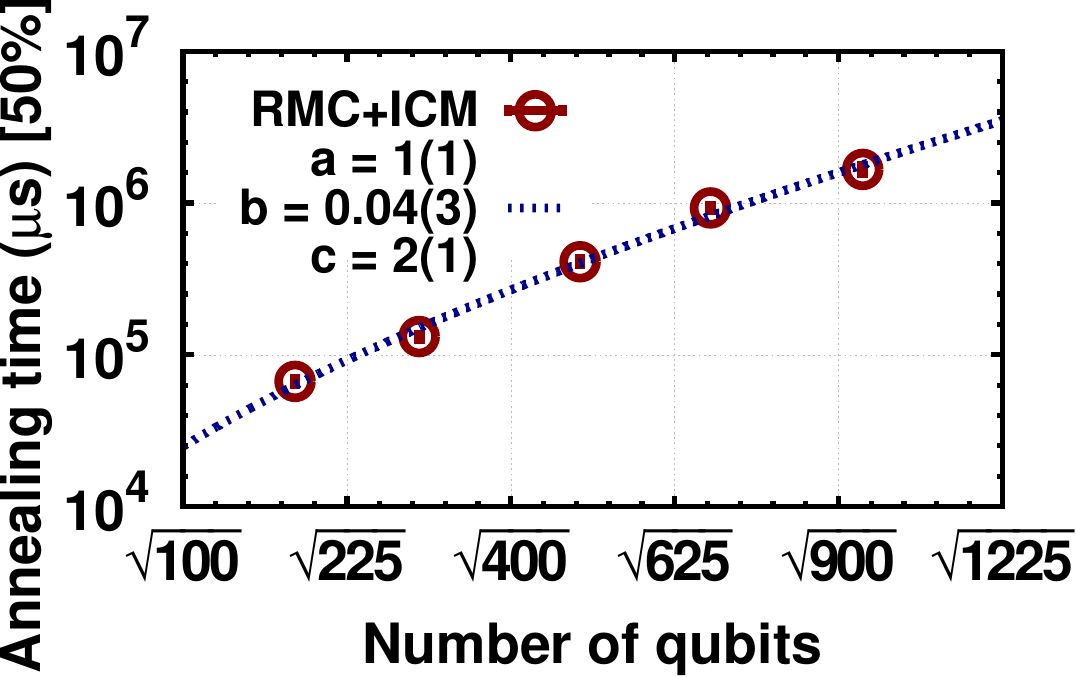}\\
\includegraphics[width=0.3\textwidth]{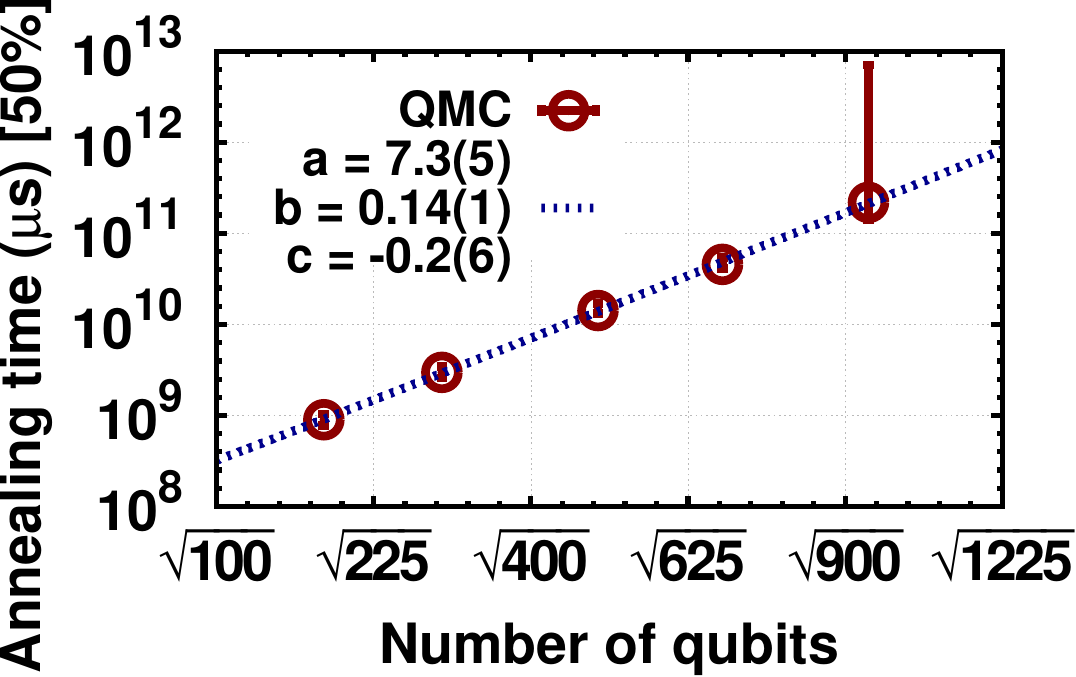}
\includegraphics[width=0.3\textwidth]{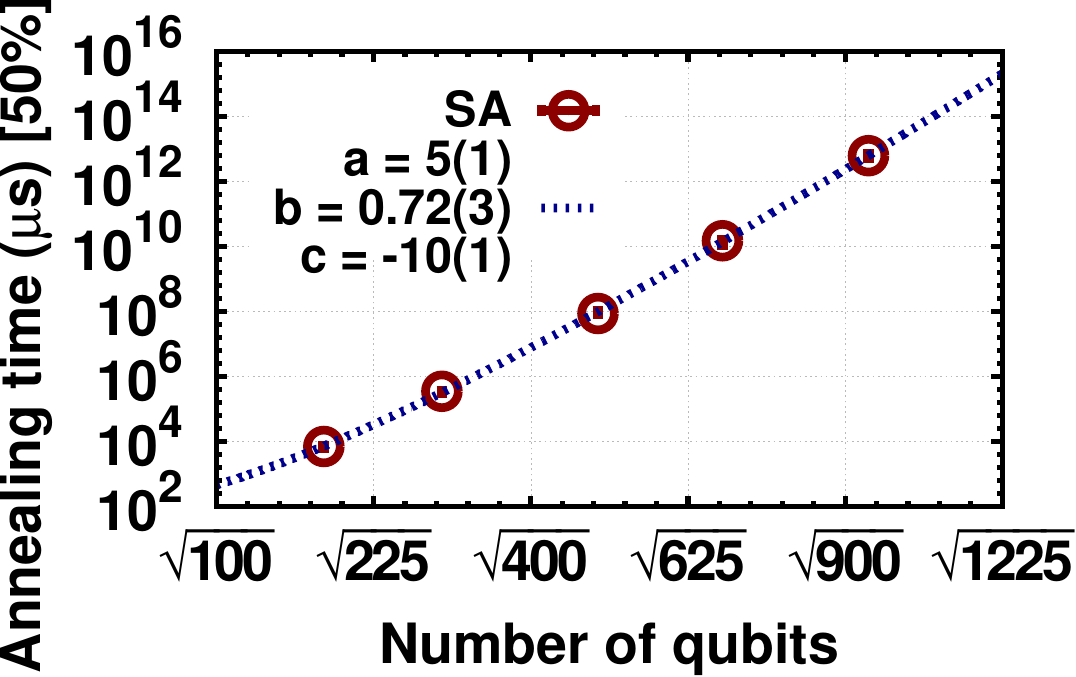}
\includegraphics[width=0.3\textwidth]{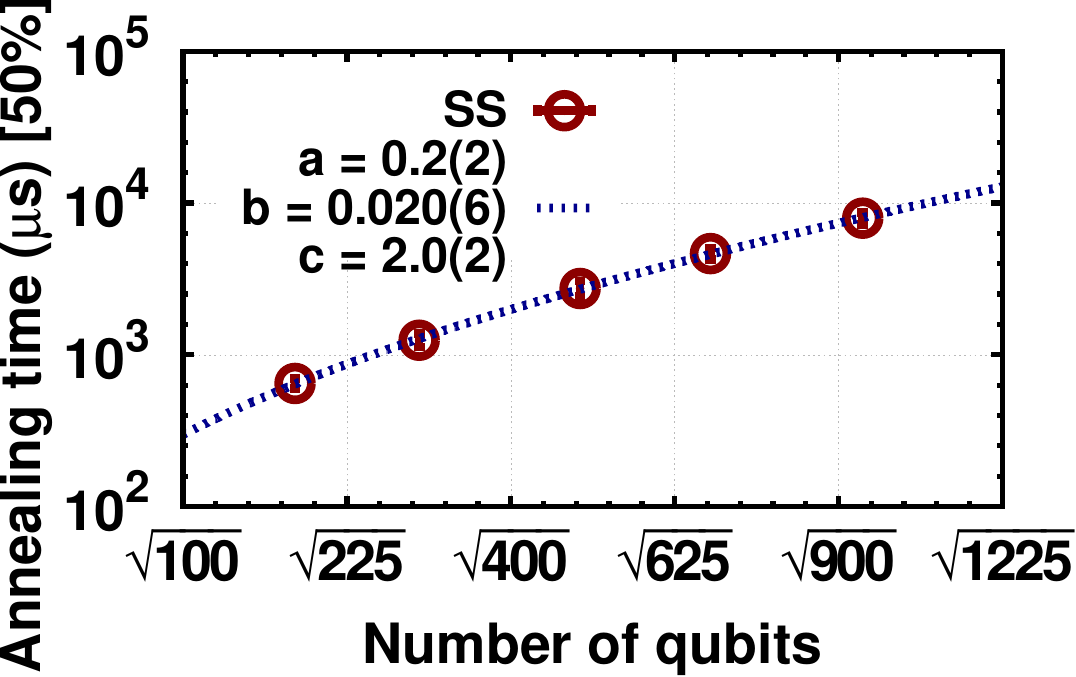}
\caption{\label{fig:fit_50}
Scaling analysis of $\log_{10} T_\text{ann}$ by using a log-corrected fit of the
form $f(n) = a + b \sqrt{n} + c \log_{10}(\sqrt{n})$, where $n$ is the
number of qubits. The panels show the values of the fit parameters and
how well the function $f(n)$ fits the data.}
\end{figure*}

\begin{figure*}[t!]
\includegraphics[width=0.3\textwidth]{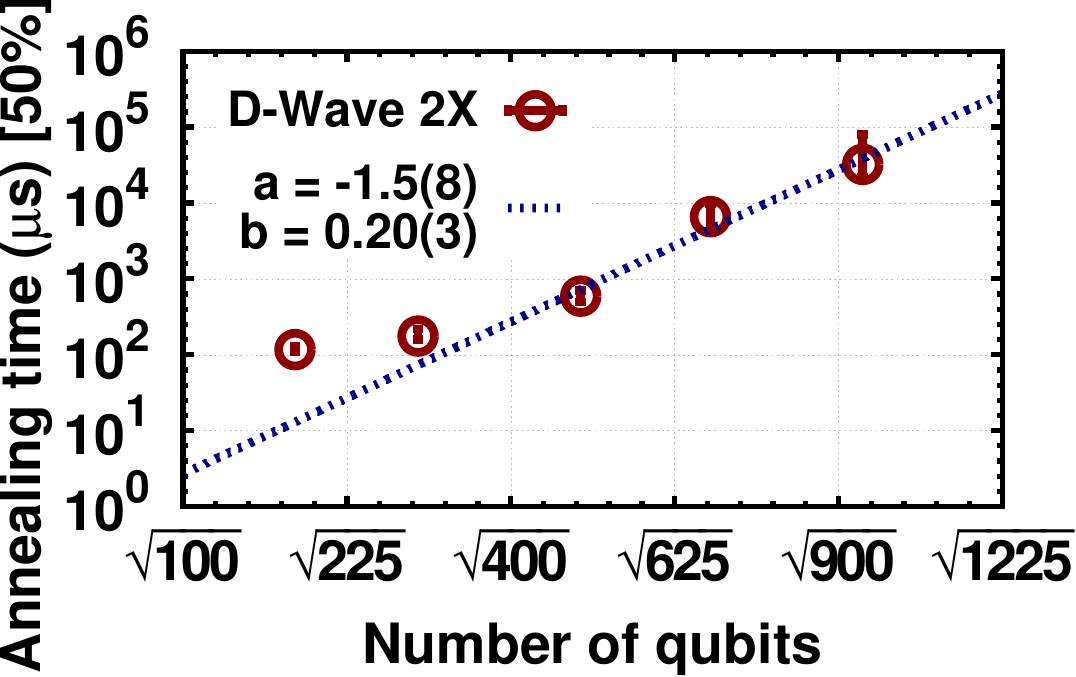}
\includegraphics[width=0.3\textwidth]{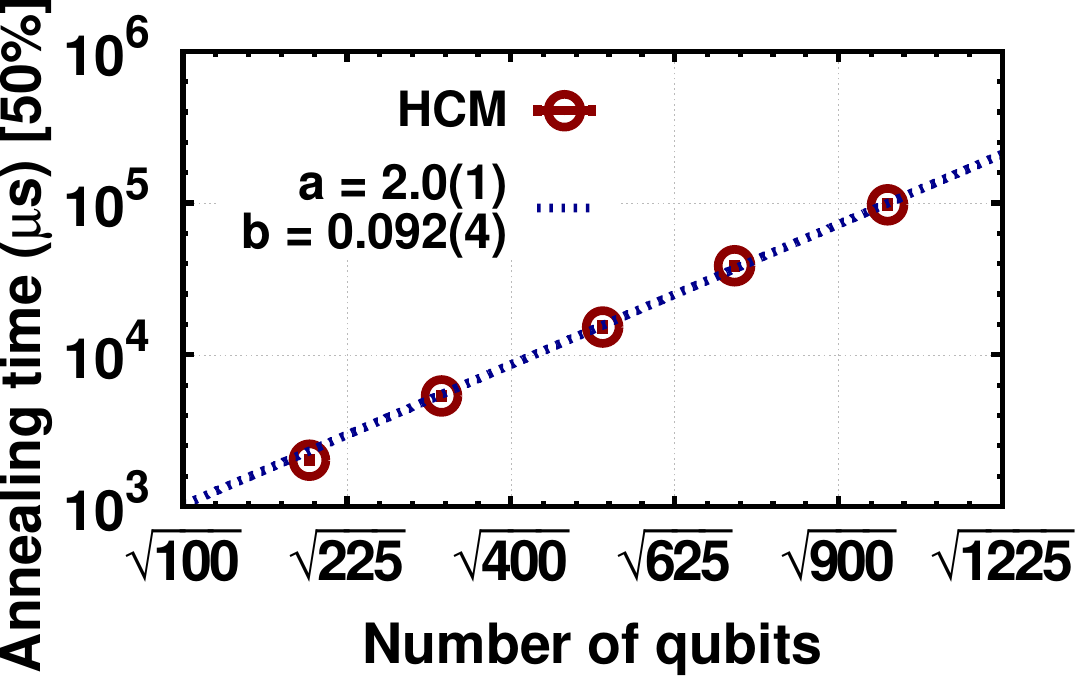}
\includegraphics[width=0.3\textwidth]{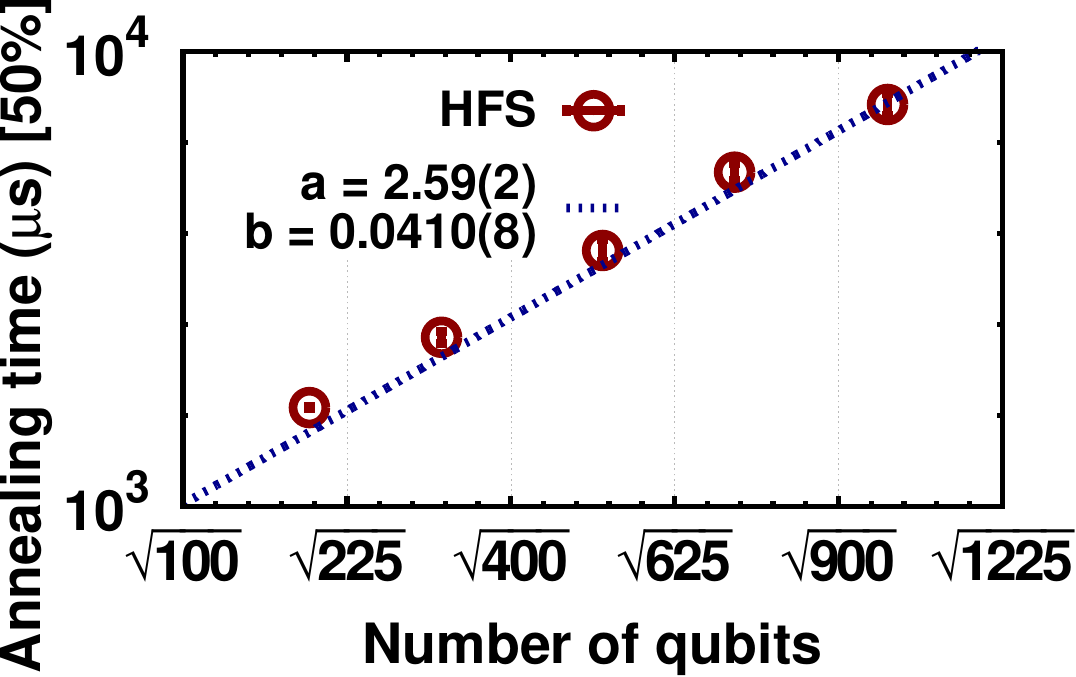}\\
\includegraphics[width=0.3\textwidth]{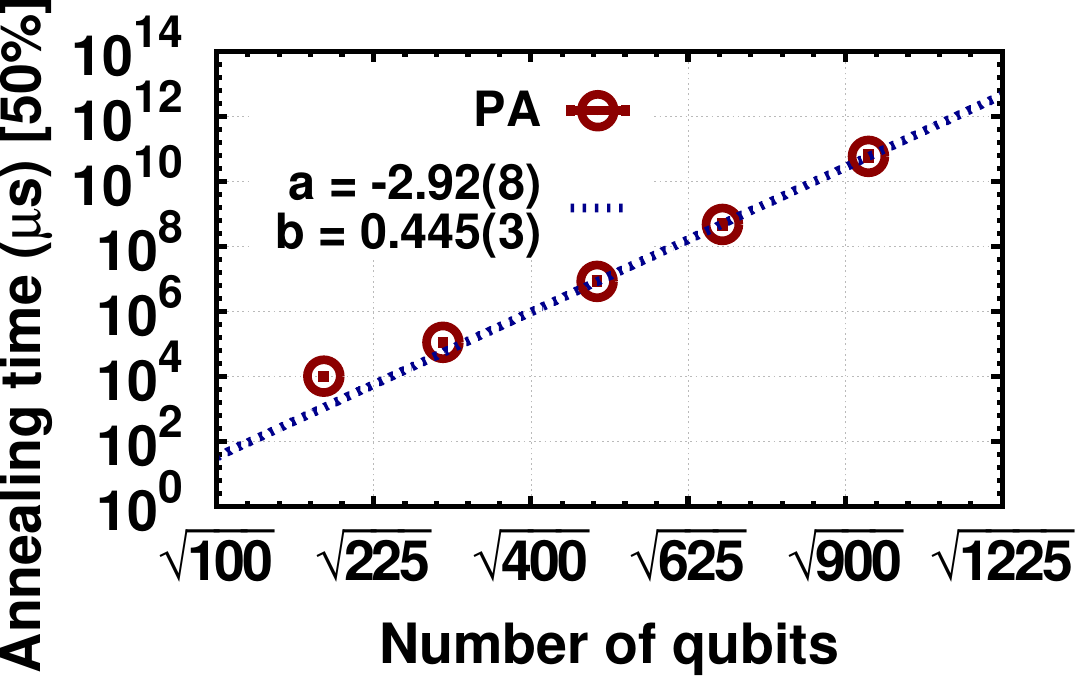}
\includegraphics[width=0.3\textwidth]{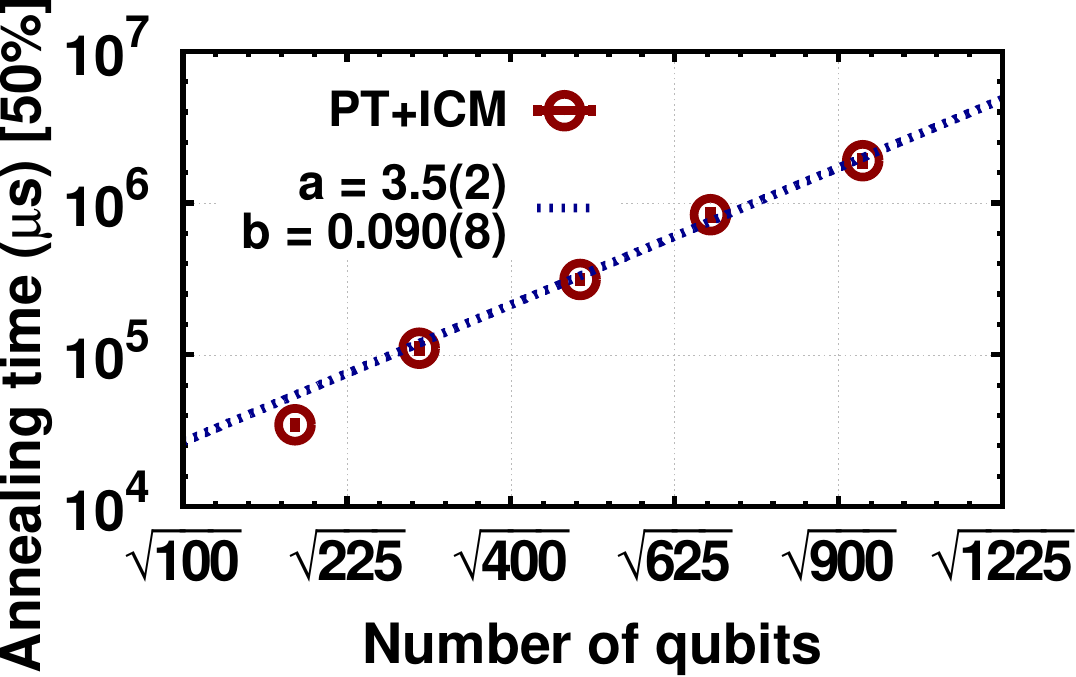}
\includegraphics[width=0.3\textwidth]{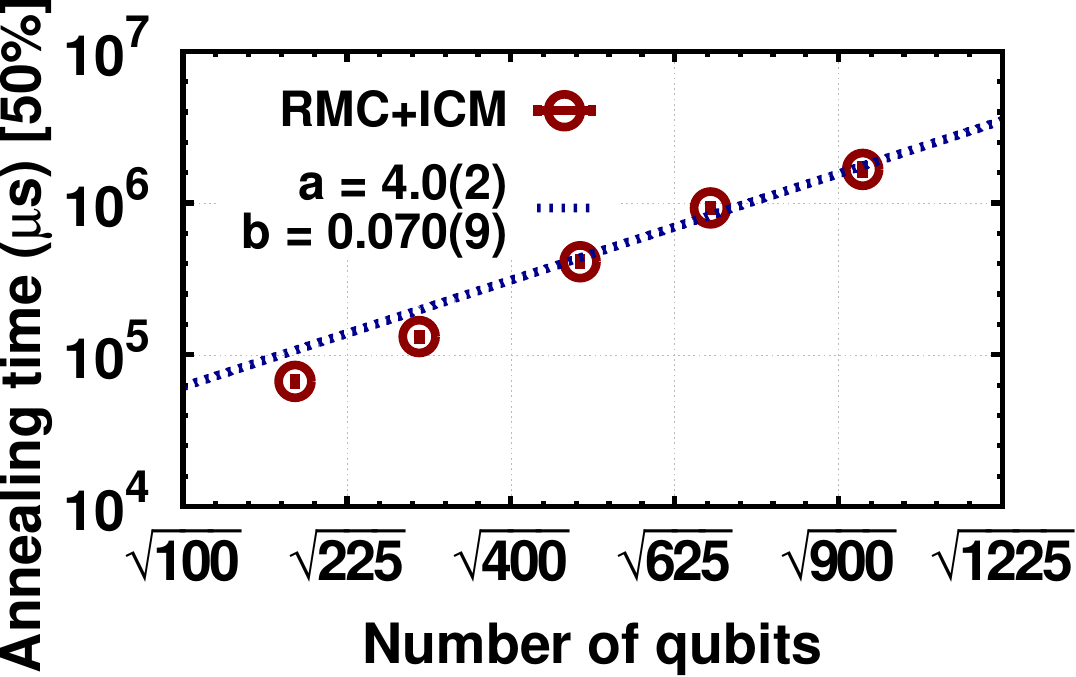}\\
\includegraphics[width=0.3\textwidth]{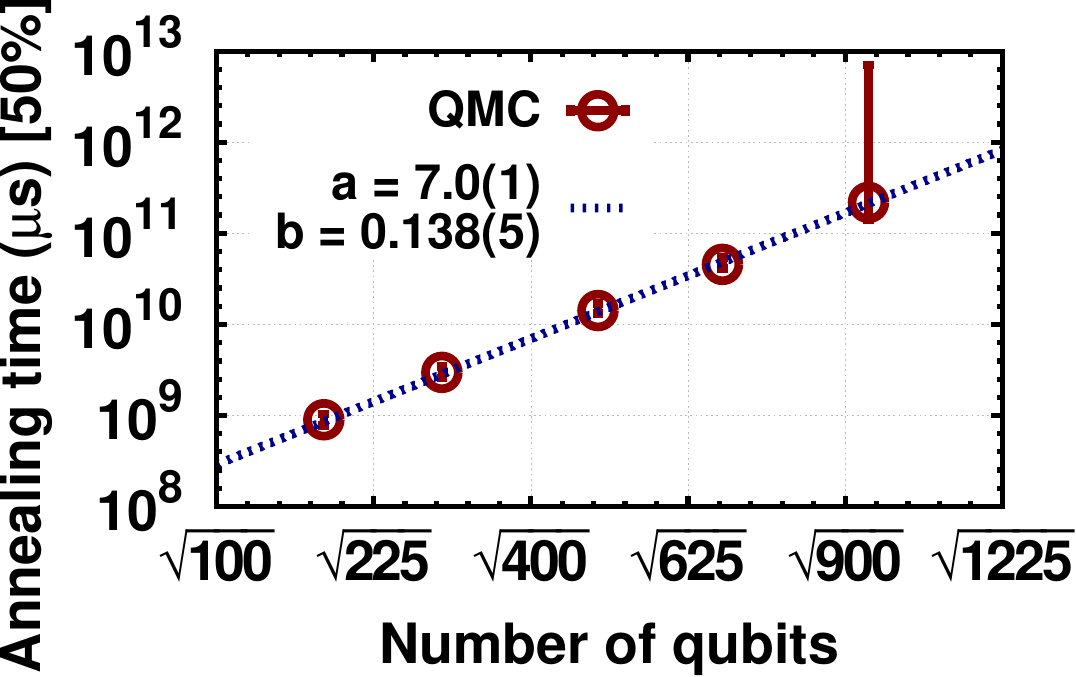}
\includegraphics[width=0.3\textwidth]{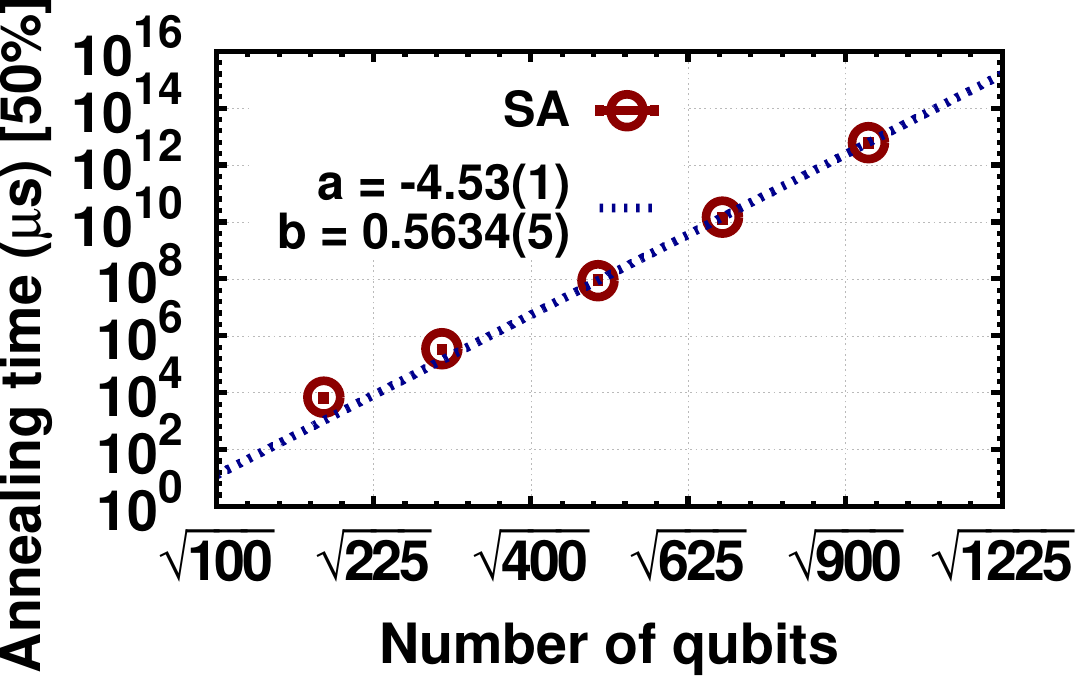}
\includegraphics[width=0.3\textwidth]{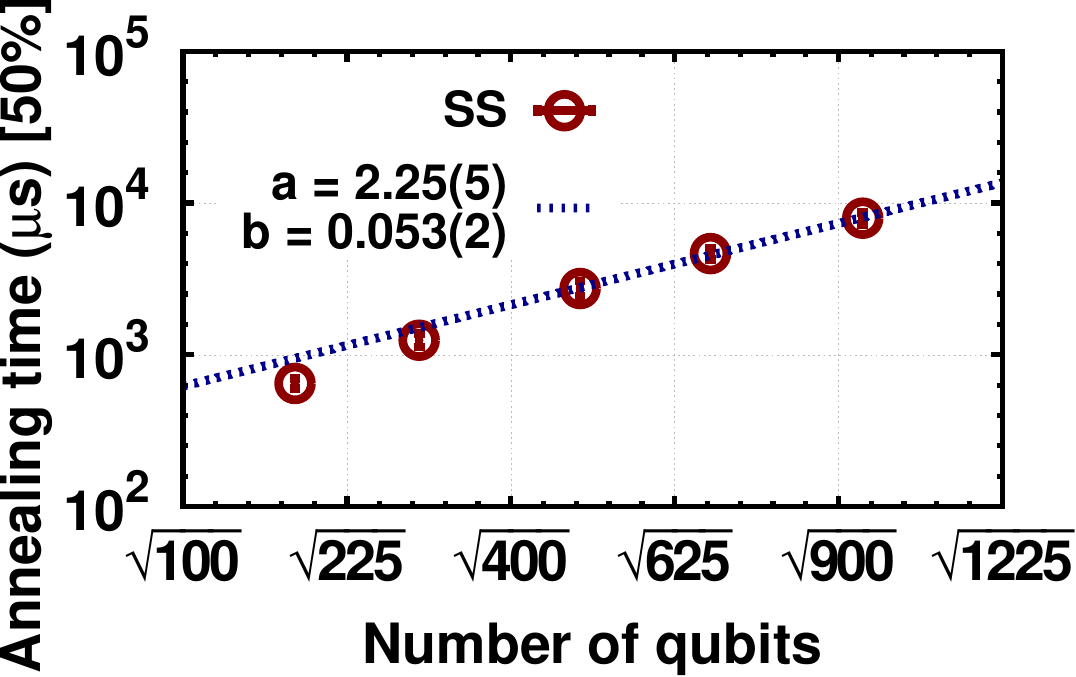}
\caption{\label{fig:fit_lin_50} Scaling analysis of
$\log_{10} T_\text{ann}$ by using a linear fit of the form
$f(n) = a + b \sqrt{n}$, where $n$ is the number of qubits. The panels
show the values of the fit parameters and how well $f(n)$ fits the data.
For the fit, only the last three data points are used.}
\end{figure*}

In the main text we define the computational time $T_\text{tts}$ for a
given classical or quantum heuristic as the time to find a solution with
$99\%$ probability \cite{ronnow:14a,boixo:14} as:
\begin{equation}\label{eq:tts_app}
	T_\text{tts} = T_\text{ann} 
	\frac{\log_{10}(1 - s)}{\log_{10}[1 - p_\text{succ}(T_\text{ann})]},
\end{equation}
where $s = 0.99$, $T_\text{ann}$ is annealing/running time and
$p_\text{succ}(T_\text{ann})$ is the probability of success at a given
$T_\text{ann}$. For the weak-strong cluster model, it is expected that
$T_\text{tts}$ will scale exponentially with the system size $n$ as
\begin{equation}
	T_\text{tts} \approx \text{poly}(n)\, 10^{a + b\,\sqrt{n}} = 10^{a + b\,\sqrt{n} + c\log_{10}\left(\sqrt{n}\right)},
\end{equation}
with $n^{c/2}$ is the dominant term of the polynomial prefactor
$\text{poly}(n)$. To determine the values of parameters $a$, $b$, and $c$
in \equationname~(\ref{eq:tts_app}) we either use a linear fit $f(x) = a
+ b\,\sqrt{n}$, i.e., it is assumed that the term $c$ is negligible, or
a log-corrected fit $f(x) = a + b\,\sqrt{n} +
c\log_{10}\left(\sqrt{n}\right)$. In \figurename~\ref{fig:scaling_50}
of the main text we report the dominant asymptotic scaling exponent $b$ of
$T_\text{tts}$ in \equationname~(\ref{eq:tts}) for the classical or quantum
heuristics presented in this paper, while in
Fig.~\ref{fig:factors_ab_50} we report the values of the parameters
$a$ and $c$. Figures \ref{fig:fit_50} and \ref{fig:fit_lin_50}
show how well either the linear regression or the
log-corrected regression fit the experimental and numerical data, respectively.

\section{Hybrid Cluster Method (HCM)}\label{sec:hcm}

The hybrid cluster method (HCM) is a Metropolis sampling technique where
``clusters'' are updated instead of single spins. The outline of HCM is
simple: given a set of connected $K$ spin-domains
$\{\mathcal{D}_i\}_{i=1,\,\ldots,\,K}$ such that their union is the
whole system, clusters are created using the Wolff rule \cite{wolff:89}
inside a randomly-chosen domain $\mathcal{D}_i$. Then the cluster is
flipped by following a Metropolis updated by considering only couplings
outside the selected domain (see \referencename~\cite{venturelli:15a}
for more details). HCM was initially developed to improve the
thermalization of highly-structured systems such as embedded systems
because it preserves the detailed balance \cite{venturelli:15a}
condition. Additionally, HCM can be used as a random heuristics for
finding ground states efficiently.

In the weak-strong cluster model, domains $\mathcal{D}_i$ are identified
as unit cells of the chimera graph. Because spins inside unit cells are
ferromagnetically coupled, they are likely to act as a single cluster in
the low-temperature regime. The system is therefore started from an
initial high temperature $1/T_\text{ini} = \beta_\text{ini} = 0.5$ and
then cooled to the final temperature $1/T_\text{end} = \beta_\text{end}
= 3$. For the simulations, a linear schedule (in the inverse temperature
$\beta$) with $M$ steps has been used, where $M$ is optimized by
minimizing Eq.~(\ref{eq:tts_app}).  At each step, a full
update of the system is made.

Table \ref{table:par_hcm} lists the simulation parameters used to compute the 
time to solution in Fig.~\ref{fig:scaling_50} of the main text.

\begin{table}[h!]
\caption{\label{table:par_hcm}Parameters of the
	simulation using the hybrid cluster method (HCM) on the chimera
	topology. $M$ represents the optimal number of inverse temperature
    steps for the thermal annealing.}
	\begin{tabular*}{\columnwidth}{@{\extracolsep{\fill}}lccc}
		\hline
		\hline
		\textbf{System size} (\boldmath{$n$}) & \boldmath{$\beta_\text{ini}$} & \boldmath{$\beta_\text{end}$} & \boldmath{$M$}\\
		\hline
		$192$ & $0.5$ & $3$ & $5$\\
		$300$ & $0.5$ & $3$ & $6$\\
		$520$ & $0.5$ & $3$ & $8$\\
		$720$ & $0.5$ & $3$ & $11$\\
		$992$ & $0.5$ & $3$ & $14$\\
		\hline
        \hline
	\end{tabular*}
\end{table}

\section{Isoenergetic Cluster Algorithm (ICM)}
\label{sec:icm}

The isoenergetic cluster method (ICM) is a rejection-free cluster
algorithm for spin glasses that greatly improves thermalization
\cite{zhu:15b}.  The main idea of ICM consists in restricting Houdayer
cluster moves \cite{houdayer:01} to temperatures where cluster
percolation is hampered by the interplay of frustration and temperature.
As such, one is able to extend the Houdayer cluster algorithm from
two-dimensional spin glasses (for which the Houdayer algorithm was
originally designed for) to any topology and/or space dimension. More
precisely, $M$ copies of the system are run at the same temperature. The
$q$-space intersection between two random replicas $\alpha$ and $\beta$
is then defined as $q_j=\sigma^{z,\alpha}_j \sigma^{z,\beta}_j$
\cite{zhu:15b}.  Within the overlap space ($q$-space) the system has two
domains: sites with $q_j = 1$ and the sites with $q_j =-1$. In ICM,
clusters are defined as the connected parts of these domains.  Once the
clusters are created, a random site with $q_j = -1$ is chosen and the
corresponding cluster flipped. Because the total energy of the two
copies of the system is unchanged by this transformation, the acceptance
of the cluster move is rejection free.  One of the main advantages of
ICM is that it allows for a more extensive exploration of the energy
landscape by classically teleporting across energy barriers. Note that
the cluster updates obey detailed balance and are only ergodic after
being combined with Monte Carlo lattice sweeps. The method is used to
improve the sampling of parallel tempering Monte Carlo (PT)
\cite{hukushima:96,katzgraber:04d,katzgraber:09e} which is the current
state-of-the-art simulation method for spin glasses.

Although the aforementioned approach is designed to quickly thermalize a
frustrated system at finite temperatures, the method can be adjusted to
act as a heuristic to find ground state configurations
\cite{katzgraber:03,moreno:03} (PT+ICM). To do this, the lowest
temperature of the simulation is chosen low enough such that the
different copies of the system at different temperatures occasionally dip
into the ground state. To verify whether the true ground state has been
reached, two criteria are adopted: first, the same minimum-energy state
has to be reached from four replicas at the minimum temperature $T_{\rm
min}$.  Second, this state has to be reached during the first $25$\% of
the sweeps in all four copies. These conditions are satisfied for the
parameters listed in Table \ref{table:par_icm}.

We have also combined ICM with the replica Monte Carlo algorithm
(RMC+ICM) \cite{wang:05a}.  The RMC algorithm is based on three basic
steps: first, $R$ replicas of the system are run at different
temperatures $\{T_1,T_2,...,T_{R}\}$. Second, a site is picked at random
and the associated cluster (which is defined through the overlap of the
systems at nearby temperatures) is created.  Third, a Metropolis update
is performed to flip the cluster.  Replica Monte Carlo is extremely
efficient in two-dimensional or quasi-two-dimensional spin glasses,
reducing the correlation time enormously compared to single spin flips
\cite{wang:05a}.  However, in higher space dimensions, its performance
is comparable to parallel tempering Monte Carlo. The parameters for the
simulations are reported in \tablename~\ref{table:par_icm}.

\begin{table}[h!]
	\caption{\label{table:par_icm}Parameters of the
	simulation using the isoenergetic cluster method (ICM) on the
	chimera topology.  $T_\text{min}$ [$T_\text{max}$] is the
	lowest [highest] temperature simulated, and $N_T$ is the total
	number of temperatures used in the parallel tempering and
	replica Monte Carlo methods. Isoenergetic cluster moves only
	occur for the lowest $N_\text{c}$ temperatures simulated.}
	\begin{tabular*}{\columnwidth}{@{\extracolsep{\fill}}ccccc}
		\hline
		\hline
		\textbf{System size} (\boldmath{$n$}) & \boldmath{$T_\text{min}$} & \boldmath{$T_\text{max}$} & \boldmath{$N_{T}$} & \boldmath{$N_\text{c}$}  \\
		\hline
		$180,\,296,\,489,\,681,\,945$ & $0.2279$ & $2.5000$ & $21$ &$5$ \\
		\hline
        \hline
	\end{tabular*}
\end{table}

\section{Population Annealing Monte Carlo (PA)}
\label{sec:pa}

Population annealing (PA) Monte Carlo is a sequential Monte Carlo
algorithm to compute equilibrium states of systems with rugged energy
landscapes \cite{hukushima:03,machta:10,machta:11,wang:15,wang:15e}.  PA
is closely related to simulated annealing in that the system is prepared
at a high temperature and then annealed to a low target temperature.
However, instead of simulating one system, in population annealing $R$
copies of the system are simulated in parallel. At each temperature step
the population of replicas is resampled such that they represent (at any
temperature) a faithful Boltzmann distribution for that given
temperature.  Once the replicas have been resampled, replicas are
updated using Metropolis sampling. In Ref.~\cite{wang:15} it
was shown that PA can be used as an optimization heuristic that clearly
outperforms simulated annealing.

\begin{table}[h!]
	\caption{\label{table:par_pa}Simulation parameters for
	population annealing Monte Carlo (PA): number of spins $n$, 
    working population size $R$, number of temperatures $N_T$, and
    number of independent runs $M$. For all the simulations, temperatures 
    are evenly chosen in the interval $\beta = [0,\,1]$ and the number
	of sweeps applied to each replica is fixed to $N_S = 10$.}
	\begin{tabular*}{\columnwidth}{@{\extracolsep{\fill}}lrcc}
		\hline
		\hline
		\textbf{System size} (\boldmath{$n$}) & \boldmath{$R$} & \boldmath{$N_T$} & \boldmath{$M$}\\
		\hline
		$180$ & $10^2$ & $100$ & $200$\\
		$296$ & $3\cdot 10^2$ & $100$ & $200$\\
		$489$ & $10^4$ & $200$ & $200$\\
		$681$ & $10^5$ & $300$ & $200$\\
		$945$ & $3\cdot 10^6$ & $300$ & $\phantom{2}55$\\
		\hline
        \hline
	\end{tabular*}
\end{table}

In the actual simulations we simulate each problem at a working
population of size $R$ and measure the success probability $p$ to find
the ground states via $M$ independent runs. The probability $p$ is then
used to calculate the critical population size $R_c$ for a $99\%$
success probability as $R_c= R\, \ln(0.01)/\ln(1-p)$. This can be
further transformed to the amount of work in Monte Carlo lattice sweeps,
and thus a physical time. Here, we use $N_T$ temperatures evenly
distributed in $\beta = 1/T \in [0,\, 1]$, and at each temperature, $N_S
= 10$ Monte Carlo sweeps are applied to each replica.
Table~\ref{table:par_pa} lists the parameters of the simulation.

\section{Super-spin heuristic (SS)}\label{sec:ss}

The weak-strong cluster model introduced in
\referencename~\cite{denchev:16} is a highly-structured problem. In
particular, \kff unit cells of the chimera graph are ferromagnetically
coupled and biased by a strong external field. Hence, spins belonging to
the same unit cell are likely to be aligned in the ground state. The
super-spin (SS) approach takes advantage of the structure of the
weak-strong clusters by identifying a single \kff cell as a
``super-spin''. The resulting ``logical'' model is therefore a
considerably smaller two-dimensional spin-glass problems with external
fields. Each of these logical spins is then coupled to an external local
field. For the example shown in \figurename~\ref{fig:weak-strong}, the
original problem size of $n = 224$ spins is reduced to a spin-glass
problem of only $224/8 = 28$ spins that is trivial to optimize.

The time-to-solution of the SS approximation is then computed by
applying the ICM+PT heuristic introduced in Appendix~\ref{sec:icm}.
Because the SS approximation does not take into account the detailed
structure of the strong-weak clusters, it is expected to be the fastest
among the different heuristics used. Indeed, as shown in
Fig.~\ref{fig:scaling_50}, results using SS are not only the
fastest, but also represent the algorithm with the best computational
scaling.

\section{Other algorithms used (QMC, SA, and HFS)}
\label{sec:otheralg}

For details on the quantum Monte Carlo (QMC) and simulated annealing
(SA) results, simulation parameters and algorithmic details we refer the
reader to Ref.~\cite{denchev:16}.  The Hamze--de-Freitas--Selby (HFS)
algorithm \cite{hamze:04,selby:14} is explained in detail in
Ref.~\cite{comment:selby}.

\section{Two-energy level system}
\label{sec:2LV}

The calculation of the probability of success
$p_\text{succ}(n,\,T_\text{ann})$ for the two-energy level model in
Eq.~(\ref{eq:ham_2LV}) has been done by a numerical
integration of the Schr\"odinger equation using a a non-optimal (linear)
schedule with a total annealing time of $T_\text{ann} = 500$. For the
integration, we have discretized the time using $\delta t = 0.01$ for
all the system sizes $n = {1,\,2,\,\ldots,\,16}$.  The time
discretization has been chosen so that there were no appreciable changes
in $p_\text{succ}(n,\,T_\text{ann})$ by decreasing $\delta t$.
Table \ref{table:par_2LV} reports the parameters used for the
two-energy level model.

\begin{table}[h!]
	\caption{\label{table:par_2LV}Parameters used for the numerical
	integration of the Schr\"odinger equation of the two-energy
	level model in \equationname~(\ref{eq:ham_2LV}).}
	\begin{tabular*}{\columnwidth}{@{\extracolsep{\fill}}cccc}
		\hline
		\hline
		\textbf{System size} (\boldmath{$n$}) & \textbf{Schedule} & \boldmath{$T_\text{max}$} & \boldmath{$\delta t$} \\
		\hline
		$1,\,2,\,\ldots,\,16$ & linear & $500$ & $0.01$ \\
		\hline
        \hline
	\end{tabular*}
\end{table}

\bibliography{refs.bib,comments.bib}

\begin{thebibliography}{88}
\bibitem[{com({\natexlab{b}})}]{comment:relaxed}
\bibinfo{note}{Note that the ``slight'' advantage found in
  Ref.~\cite{katzgraber:15} for the D-Wave Two quantum annealer over classical
  simulated annealing has to be seen in a relative sense. While the D-Wave Two
  quantum annealer failed to solve most problems, it performed better for
  problems with thinner energy barriers. Furthermore, if the criterion for
  success is relaxed to include the $k$ lowest energy states, the performance
  of the D-Wave Two improved noticeably in comparison to classical simulated
  annealing.}
\bibitem[{com({\natexlab{c}})}]{comment:analogy}
\bibinfo{note}{A simple analogy is the following: Suppose the reader wants to
  compare two motorbikes. Surely, it would make no sense to race a dirt bike
  against a superbike on a dirt road, because the slicks on the superbike would
  have no traction on dirt. Similarly, racing the dirt bike against a superbike
  on a paved racetrack would be futile, because the dirt bike would have little
  to no grip on the pavement. Therefore, it is important to introduce
  categories where the individual types of motorbikes are compared, i.e.,
  motocross and the superbike series.}
\bibitem[{com({\natexlab{d}})}]{comment:firas}
\bibinfo{note}{Note that some algorithms might share elements from multiple
  categories. In this case, it will be difficult to define the type of speedup.
  However, one should keep in mind that the goal is to make the fairest
  comparison possible.}
\bibitem[{com({\natexlab{f}})}]{comment:sap}
\bibinfo{note}{Simulations using the groups approach \cite{zintchenko:15} show
  a similar scaling to HCM and ICM, but are slower than the HFS and SS
  algorithms.}
\end{thebibliography}

\end{document}